\documentclass[manuscript,screen]{acmart}
\usepackage{longtable}
\usepackage{enumitem}
\usepackage{float}
\usepackage{xcolor}

\AtBeginDocument{%
  }

\setcopyright{acmlicensed}
\copyrightyear{2018}
\acmYear{2018}
\acmDOI{XXXXXXX.XXXXXXX}
\acmConference[Conference acronym 'XX]{Make sure to enter the correct
  conference title from your rights confirmation email}{June 03--05,
  2018}{Woodstock, NY}
\acmISBN{978-1-4503-XXXX-X/2018/06}

\begin{document}


\title{Relational Mediators: LLM Chatbots as Boundary Objects in Psychotherapy}

\author{Jiatao Quan}
\orcid{0009-0009-1205-769X}
\authornote{Both authors contributed equally to this research.}
\email{jq36@uw.edu}
\affiliation{%
  \institution{The Hong Kong Polytechnic University}
  \city{Hong Kong}
  \country{China}
}

\author{Ziyue Li}
\orcid{0009-0000-5628-8718}
\authornotemark[1]
\email{liz225@uw.edu}
\affiliation{%
  \institution{University of Washington}
  \city{Seattle}
  \state{WA}
  \country{USA}
}

\author{Tian Qi Zhu}
\orcid{0000-0003-0071-4081}
\email{azhu98@uw.edu}
\affiliation{%
  \institution{University of Washington}
  \city{Seattle}
  \state{WA}
  \country{USA}
}

\author{Yuxuan Li}
\orcid{0009-0002-9408-7524}
\email{lyx2004@uw.edu}
\affiliation{%
  \institution{University of Washington}
  \city{Seattle}
  \state{WA}
  \country{USA}
}

\author{Baoying Wang}
\orcid{0009-0000-9154-7367}
\email{ednawang28@gmail.com}
\affiliation{%
  \institution{University of Washington}
  \city{Seattle}
  \state{WA}
  \country{USA}
}

\author{Wanda Pratt}
\email{wpratt@uw.edu}
\orcid{0000-0003-4035-0198}
\affiliation{%
  \institution{University of Washington}
  \city{WA}
  \country{USA}
}

\author{Nan Gao}
\email{nan.gao@nankai.edu.cn}
\orcid{0000-0002-9694-2689}
\authornote{Corresponding author.}
\affiliation{%
  \institution{Nankai University}
  \city{Tianjin}
  \country{China}
}
\renewcommand{\shortauthors}{Quan et al.}

\begin{abstract}
As large language models (LLMs) are embedded into mental health technologies, they are often framed either as tools assisting therapists or autonomous therapeutic systems. Such perspectives overlook their potential to mediate relational complexities in therapy, particularly for systemically marginalized clients. Drawing on in-depth interviews with 12 therapists and 12 marginalized clients in China, including LGBTQ+ individuals or those from other marginalized backgrounds, we identify enduring relational challenges: difficulties building trust amid institutional barriers, the burden clients carry in educating therapists about marginalized identities, and challenges sustaining authentic self-disclosure across therapy and daily life. We argue that addressing these challenges requires AI systems capable of actively mediating underlying knowledge gaps, power asymmetries, and contextual disconnects. To this end, we propose the \textbf{Dynamic Boundary Mediation Framework}, which reconceptualizes LLM-enhanced systems as adaptive boundary objects that shift mediating roles across therapeutic stages. The framework delineates three forms of mediation: \textit{Epistemic} (reducing knowledge asymmetries), \textit{Relational} (rebalancing power dynamics), and \textit{Contextual} (bridging therapy-life discontinuities). This framework offers a pathway toward designing relationally accountable AI systems that center the lived realities of marginalized users and more effectively support therapeutic relationships.
\end{abstract}

\begin{teaserfigure}
    \includegraphics[width=.95\textwidth]{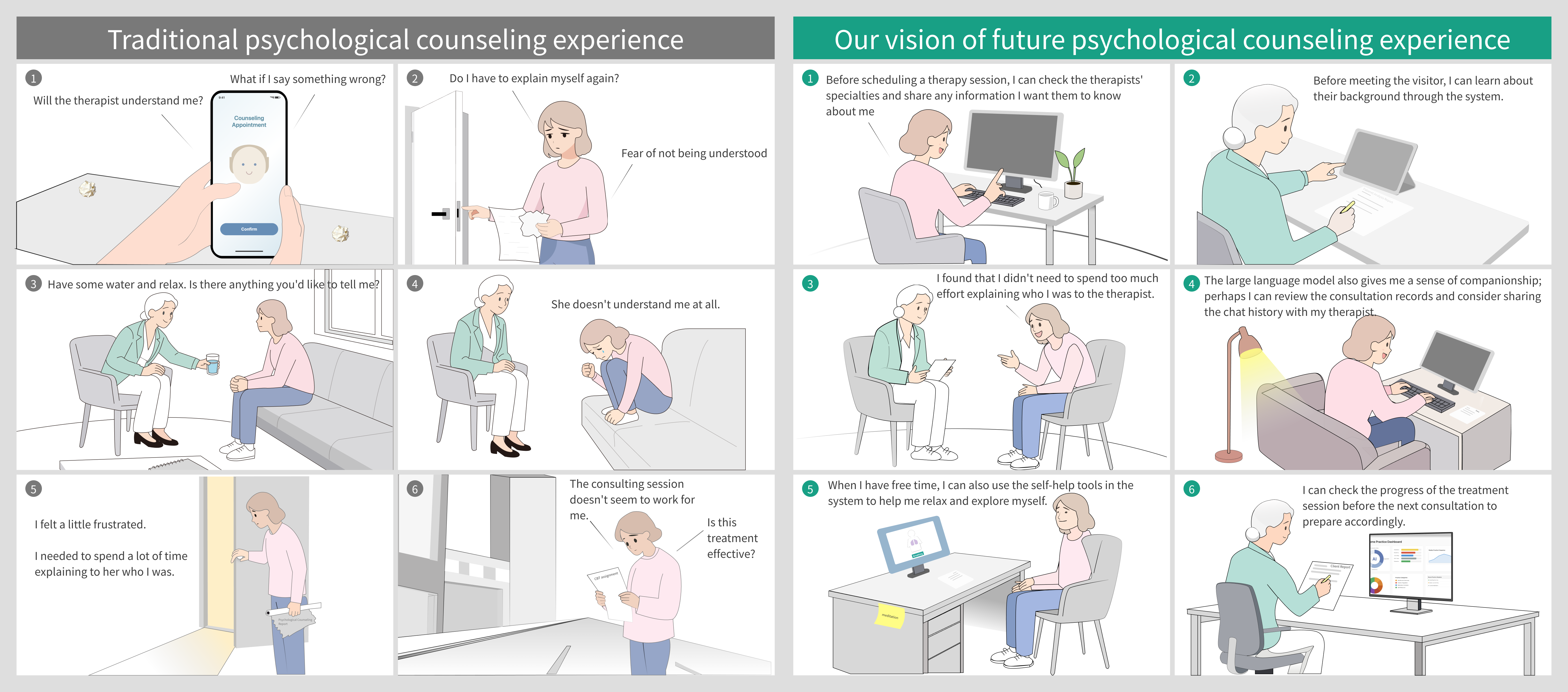}
    \caption{An illustration of how LLM chatbots can enhance the therapeutic experience as boundary objects in psychotherapy}
    \Description{Image description for accessibility.}
    \label{fig:teaser}
\end{teaserfigure}

\ccsdesc[300]{Social and professional topics~Cultural characteristics}
\ccsdesc[500]{Human-centered computing~Empirical studies in accessibility}
\ccsdesc[500]{Human-centered computing~Empirical studies in HCI}

\keywords{Human–AI Collaboration, Large Language Models, Mental Health, Boundary Objects, Psychotherapy, Marginalized Communities}

\received{20 February 2007}
\received[revised]{12 March 2009}
\received[accepted]{5 June 2009}

\maketitle

\section{Introduction}
Access to mental health care remains structurally unequal across the world, particularly for clients from marginalized communities who must navigate stigma, resource shortages and institutional barriers. In China, these challenges are especially pronounced. Despite a lifetime prevalence of depression and anxiety reaching 7.6\%~\cite{liu_prevalence_2023}, a severe shortage of qualified mental health professionals leaves more than half of the population without adequate care, with disparities even sharper for rural and under-resourced regions~\cite{xiang_rethinking_2018}. For groups such as LGBTQ+ individuals or those from low-income backgrounds, the barriers are intensified by social stigma and institutional neglect~\cite{lin_lgbtq_2022}.


In parallel, artificial intelligence (AI) is playing an increasingly prominent role in healthcare, supporting clinical decision-making, diagnostics, and treatment planning~\cite{auf_use_2025,olawade_enhancing_2024}. In the mental health domain, recent studies have begun to explore the application of large language models (LLMs)~\cite{filienko_toward_2024}. However, most existing systems focus on self-help or automated content delivery and operate independently of therapist-client relationships. As a result, they give limited attention to the interpersonal processes that make psychotherapy effective, including building the trust, developing mutual understanding and sustaining emotional resonance. This relational dynamics are especially vulnerable in contexts marked by cultural mismatch, social stigma or institutional power asymmetries. Addressing these challenges requires rethinking the role of AI not simply as a content generator, but as a collaborator in therapeutic interaction.


To advance this rethinking, we draw on \textit{Boundary-Object Theory}~\cite{star_institutional_1989} and conceptualize LLM-enhanced collaborative mental health systems as \textbf{relational mediators}. Rather than functioning as static tools for information delivery, such systems can facilitate therapist-client communication, support paced self-disclosure, and create space for emotional processing. Their role may shift across therapeutic contexts. For example, they can assist in translating unarticulated emotions into words, clarifying misunderstandings, or sustaining relational continuity between sessions. This perspective positions AI systems not only as technical artifacts but as sociotechnical actors with the potential to shape therapeutic relationships.

To empirically examine this potential, we conducted semi-structured interviews with 12 psychotherapists and 12 marginalized clients raised in Chinese cultural context. Our study is guided by two research questions:
\begin{enumerate}
    \item \textit{What challenges and dynamics shape trust, self-disclosure, and communication
between therapists and clients from marginalized groups throughout the therapeutic stages?}
    \item \textit{How can LLM-enhanced collaborative mental health systems support and improve therapy
quality across therapeutic stages?}
\end{enumerate}

We present a five-stage model of therapeutic interaction and examine how LLM-supported systems might adopt different mediating roles across these stages. This analysis reveals the potential for AI tools to help bridge divides in contexts where trust is fragile and therapist-client alignment is difficult to achieve is difficult to achieve. Our contributions span the \textbf{theoretical, empirical, and design dimensions}:

\begin{itemize}
    
\item \textbf{Theoretical Contribution.} We introduce the \textit{Dynamic Boundary Mediation Framework}, which reconceptualizes LLM-enhanced mental health systems as adaptive boundary objects that shift mediating roles across therapeutic stages. The framework outlines three forms of mediation: \textit{Epistemic} (reducing knowledge asymmetries), \textit{Relational} (rebalancing power dynamics), and \textit{Contextual} (bridging therapy-life gaps).

\item \textbf{Empirical Contribution.} Through interviews with 12 therapists and 12 marginalized clients in China, we identify sociotechnical mechanisms through which LLM-enhanced systems may mediate therapist-client asymmetries in trust-deficient contexts. Our five-stage boundary negotiation model illustrates how relational challenges manifest differently across therapeutic phases.

\item \textbf{Design Contribution.} We derive five actionable design guidelines for developing relationally accountable AI systems that center marginalized users' lived realities. These guidelines address stage-aware role shifting, negotiable data visibility, contextualized relational memory, community-validated onboarding, and pervasive context-adaptive empathy.
\end{itemize}

Although situated within mental health, our work offers broader implications for AI-mediated collaboration in other high-stakes domains such as education and social services, where asymmetrical expertise, institutional mistrust, and relational fragility shape user experiences.
\section{Related Work}

\subsection{Collaboration Challenges in Technology-Supported Mental Health}

Trust, communication and paced self-disclosure are central to effective psychotherapy across therapeutic orientations. Rogers' classic conditions for therapeutic change (e.g., empathy, unconditional positive regard and congruence) \cite{rogers_necessary_1957} and Bordin's tripartite model of the working alliance ~\cite{bordin_generalizability_1979} underscore the centrality of emotional bonding, goal consensus and mutual understanding \cite{horvath_alliance_2011}. Self-disclosure, in particular, enables clients to articulate inner experiences and deepen relational engagement~\cite{jourard_self-disclosure_1971}. For marginalized populations, however, these dynamics become more complex: minority stress~\cite{meyer_prejudice_2003} and culturally rooted mistrust in systemic oppression~\cite{sue_counseling_1999} can inhibit disclosure and impede the formation of safety in therapeutic relationships. These longstanding challenges underscore the importance of understanding how technology-mediated interventions might support or complicate core therapeutic processes.

Digital mental health tools have evolved from static support systems to more interactive and dialogical platforms. Early systems were often based on structured interventions like cognitive behavioral therapy (CBT), which provided predetermined content flows that limited adaptability to individual needs~\cite{mohr_behavioral_2010}. Although the effectiveness of CBT-based self-help tools has been demonstrated, the potential negative consequences of Internet-delivered CBT (ICBT) have often been overlooked~\cite{andersson_advantages_2014}. These tool-centric models emphasized behavior modification or symptom monitoring, but largely neglected the relationship dynamics that define therapeutic collaboration.

With the advancement of natural language processing, AI-based self-help systems have begun to support more conversational and adaptive interactions. Early examples like Woebot and Wysa employ structured CBT principles and natural conversation patterns to deliver support. Woebot, for instance, models its dialogue on clinical decision-making processes and social discourse dynamics, and has been shown to significantly reduce depressive symptoms in a comparative study~\cite{fitzpatrick_delivering_2017}. Wysa similarly draws on evidence-based techniques to promote self-awareness and personal resilience~\cite{sinha_understanding_2023}. These systems mark a shift from unidirectional guidance toward bidirectional dialogue, laying the groundwork for more relational forms of engagement between clients and therapists.

Building on this trajectory, the emergence of large language models (LLMs) such as ChatGPT and LLaMA has further expanded the interactional capabilities of digital mental health tools. Unlike earlier rule-based or template-driven agents, LLM-driven systems can interpret context, detect subtle emotional cues~\cite{fu_laerc-s_2025}, and generate nuanced, personalized responses~\cite{maity_future_2024}, enabling AI to serve as a potential relational mediator. However, most existing research focuses on performance, usability, or engagement outcomes, often overlooking how these tools mediate collaborative processes~\cite{lawrence_opportunities_2024}.

\textbf{Relational Breakdowns and Trust Challenges in Digital Therapy.} Despite these technological advances, digital mental health interventions face significant relational challenges that extend beyond technical capabilities. Research has documented various forms of relationship breakdowns in technology-supported therapy, including trust ruptures, cultural misalignments, and power asymmetries that can undermine therapeutic effectiveness~\cite{borghouts_barriers_2021, berardi_barriers_2024}.

Trust ruptures in digital therapy contexts often manifest through users' skepticism about AI systems' ability to
understand their unique cultural contexts and personal experiences~\cite{timmons_call_2023}. For instance, LGBTQ+ individuals may experience identity conflicts when AI systems fail to recognize or appropriately respond to their specific mental health needs, particularly given their documented higher rates of discrimination in healthcare settings and reluctance to disclose sexual orientation due to provider bias~\cite{kormilitzin_participatory_2023}. 

Cultural misalignments further complicate digital therapeutic relationships, particularly when systems are designed within Western-centric frameworks that fail to account for diverse cultural values and communication patterns~\cite{vandecasteele_barriers_2024}. In collectivist cultures, for example, the emphasis on individual self-disclosure in many digital therapy platforms may conflict with cultural norms that prioritize group harmony and collective privacy management over individual expression.~\cite{knijnenburg_cross-cultural_2022}. These misalignments can create communication barriers that prevent effective therapeutic collaboration and may even reinforce existing power asymmetries between users and healthcare systems.

\textbf{Power Asymmetries and Structural Inequalities.} Digital health systems often perpetuate existing power imbalances by positioning users as passive recipients of care rather than active collaborators in their own healing process~\cite{ogard-repal_addressing_2025}. This top-down approach can be particularly problematic for marginalized populations who already face structural barriers in accessing mental health services. Research has shown that when AI systems fail to acknowledge or address power imbalances, including manipulative behaviors or algorithmic conformity, they can inadvertently reinforce feelings of disempowerment and reduce users' willingness to engage authentically with the therapeutic process.~\cite{zhang_dark_2025}.

The intersection of these relational challenges, including trust ruptures, cultural misalignments, and power asymmetries, creates complex barriers to effective therapeutic collaboration in digital contexts. Traditional approaches to AI system design, which focus primarily on technical performance and user engagement metrics, often fail to address these deeper relational dynamics. This gap highlights the need for a more nuanced understanding of how AI systems can serve as dynamic mediators that actively work to bridge these relational divides.

Our work addresses this gap by reconceptualizing LLM-powered systems as dynamic boundary objects that can adapt to shifting relational contexts, actively mediating trust-building, cultural understanding, and power negotiation in ways that traditional digital therapy tools cannot achieve.

\subsection{Boundary Objects and Human-AI Collaboration in Sensitive Care Contexts}

The concept of boundary objects, first introduced by Star and Griesemer~\cite{star_institutional_1989}, provides a powerful theoretical lens for understanding collaboration across diverse social worlds. Boundary objects function as bridges between communities with different goals, knowledge systems, or perspectives, facilitating coordination while accommodating local variations~\cite{lee_boundary_2007,haland_care_2015}. In healthcare, the productive vagueness of the boundary object is essential because care work spans heterogeneous professional logics and inter-organizational boundaries\cite{allen_boundary_2009}, requiring objects that can mediate divergent understandings of care.

\textbf{Theoretical Positioning in Mental Health Contexts.} The application of boundary object theory to mental health interventions requires careful consideration of the unique relational dynamics that characterize therapeutic interactions. Unlike traditional boundary objects that primarily facilitate information exchange, mental health contexts resist such direct forms of exchange. Instead, the boundary object should act as an operational articulation of Winnicott’s potential space, an in-between zone in which meaning is affectively and continually negotiated through the sustained relationship between therapist and client\cite{benezer_winnicotts_2012}. Therefore mental health contexts demand boundary objects that can navigate complex emotional landscapes and facilitate meaningful engagement across different social worlds~\cite{terlouw_boundary_2022}. This complexity becomes particularly pronounced when considering cultural sensitivities, power asymmetries, and the needs of marginalized populations who face structural barriers in accessing quality care.

\textbf{Why Boundary Object Theory?} Several theoretical frameworks could potentially illuminate AI systems' role in therapeutic collaboration. For instance, mediation theory, originally articulated by Jean Gagnepain in the 1960s, emphasizes the role of symbols and cognition as processes that facilitate communication and understanding between different entities~\cite{jensen_tropisetron_2000}. While it provides valuable insights into coordination and information exchange, traditional mediation approaches face methodological limitations when dealing with complex, interactive relationships~\cite{mackinnon_correspondence_2020}. This view may overlook the dynamic, ongoing negotiation of meaning and the fluid power relations characteristic of therapeutic relationships.

Another theoretical perspective often applied to sociotechnical systems is actor–network theory (ANT). ANT conceptualizes systems as networks of human and non-human entities and adopts a relational view of action. It's also a useful lens to understand the intervention-context interactions and unpack its mechanisms and effects. However, by distributing agency across humans and non-humans, ANT risks diffusing human intentionality, ethics, and responsibilities, which are essential in psychotherapy as therapists are ethically and legally bound to uphold care standards.
Furthermore, ANT looks into how technologies can be enrolled as agentic actors that participate in stabilizing networks through processes of translation, in which actors are aligned by passing through obligatory passage points that channel diverse interests into a single discourse of certainty~\cite{callon_elements_1984}. This framing is misaligned with therapeutic practice where meaning-making is purposefully kept open-ended as the therapists and clients interpret, negotiate, and align divergent forms of knowledge. ANT risks reducing actors' roles to discrete actions, overlooking its significance as a representational artifact that is understood and used differently by those engaged in the therapeutic process.

Boundary object theory addresses these limitations by providing a more comprehensive perspective that can simultaneously handle meaning negotiation, power dynamics, and adaptive mediation. The emphasis of this framework on objects that maintain coherence while adapting to local needs \cite{star_institutional_1989}, which makes it particularly well aligns with the dynamic, context-sensitive nature of therapeutic interactions, especially when supporting marginalized populations who face unique structural challenges. Moreover, the recognition of the theory that boundary objects can operate as "both mediators in the production of knowledge and vectors of translation in the organization of heterogeneous worlds" \cite{trompette_revisiting_2009}, offers crucial insights for understanding how AI systems might function as adaptive mediators in evolving therapeutic relationships. 

Boundary object theory offers a valuable lens for examining technologies in healthcare systems, where facilitating cooperation between heterogeneous actors and organizations is essential. Hansen et al. ~\cite{hansen_beyond_2023} explored how electronic health records affected the stakeholders’ interactions in the healthcare system, focusing on the changes to field and boundary spanning practices. Saidi et al.~\cite{saidi_crossing_2023} studied a video-supported digital toolkit used between caregivers and supporters, finding it bridged the knowledge gap while enhancing engagement and emotional connections. Within mental health specifically, Sanders et al.~\cite{sanders_boundary_2021} investigated electronic mental health screener (MHS) as a boundary object to support officers’ risk assessments and facilitate collaboration with emergency departments and social services, with an emphasis on its effect on officer’s decision making and information sharing. 

While early conceptualizations of boundary objects often emphasized their role as static intermediaries facilitating information exchange~\cite{star_institutional_1989}, more recent work has challenged this view by highlighting their dynamic evolution and entanglement with power relations in interdisciplinary collaboration~\cite{filippi_interdisciplinarity_2023}. However, these discussions remain underexplored in sensitive care contexts, where emotional vulnerability and structural asymmetries may intensify the need for boundary objects to mediate not just meaning but also trust, identity, and disclosure.

Despite advances in human-computer interaction research, the application of boundary object theory to LLM-based mental health interventions remains particularly underdeveloped. Research exploring relational dimensions of AI-supported tools rarely examines how these systems might address structural inequalities faced by marginalized populations.

This gap highlights the need to reconceptualize boundary objects not as static infrastructures that merely facilitate interoperability, but as dynamic, agentive mediators that actively shape trust, self-disclosure, and communication in context-sensitive and affective ways. In particular, AI-based mental health systems offer a compelling site for examining how boundary objects may shift roles across therapeutic contexts, particularly when supporting populations facing power asymmetries or limited access to traditional care. This perspective, which we adopt and develop in this study, not only extends boundary object theory by emphasizing dynamism and relational agency but also critically informs the development of more inclusive and relationally attuned system designs.
\section{Methodology}

This section outlines our qualitative research design and analytical approach. We begin by establishing our methodological foundations and theoretical positioning, followed by participant recruitment and sampling, data collection procedures through interviews and design probes, systematic analysis, and ethical considerations.

\begin{figure}
  \centering
  \includegraphics[width=0.95\textwidth]{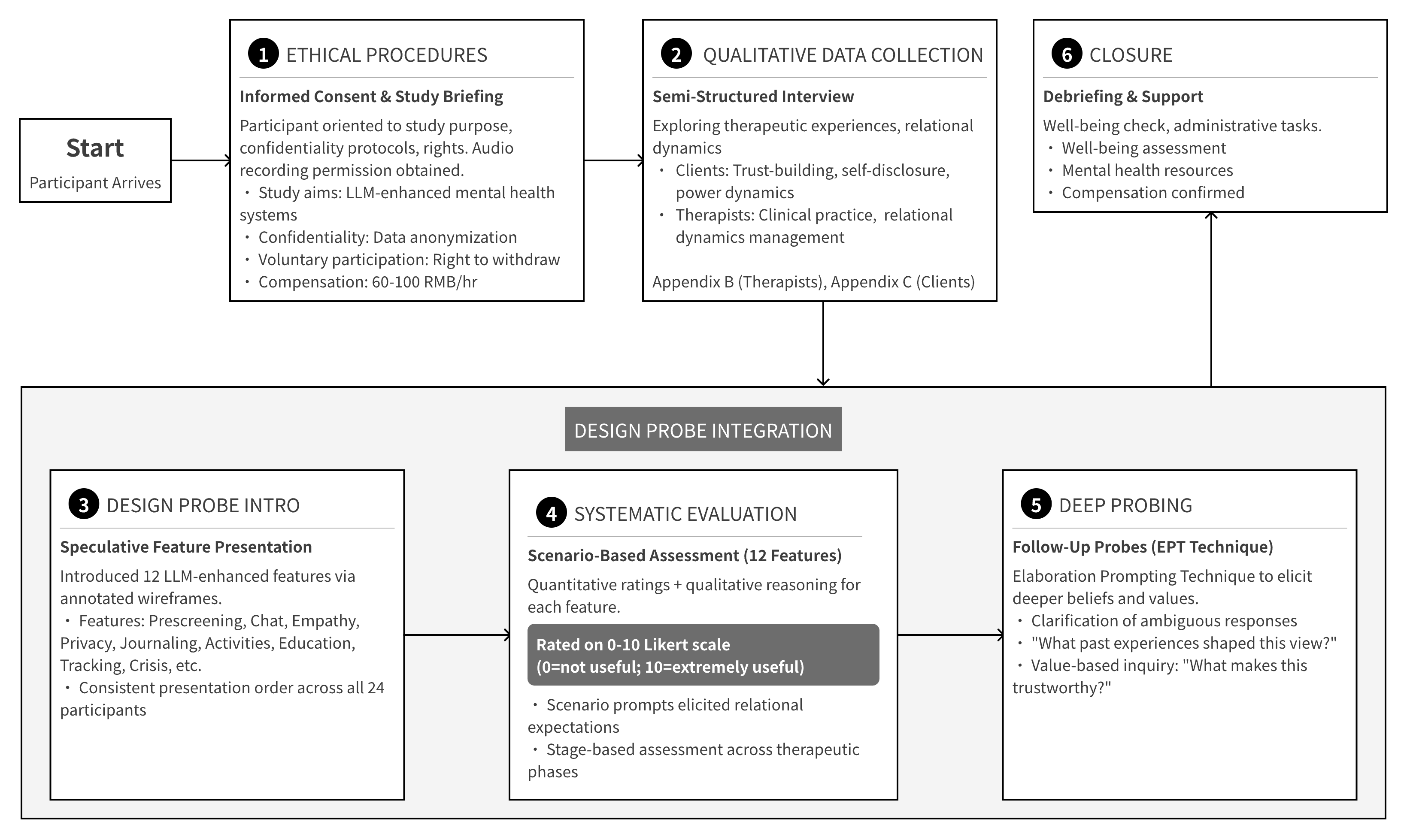}
  \caption{Overview of the research methodology workflow, illustrating the complete process from participant recruitment through data collection to analysis. The workflow integrates ethical procedures, semi-structured interviews, and design probe evaluation using the Elaboration Prompting Technique (EPT) to elicit in-depth perspectives on LLM-enhanced mental health systems.}
  \label{fig:method_overview}
\end{figure}

\subsection{Methodological Orientation and Theoretical Framework}
This study adopts a constructivist qualitative research approach to examine how AI systems function as dynamic \textit{boundary objects}~\cite{star_institutional_1989} mediating therapeutic relationships in digital mental health services. Rather than focusing solely on the functional attributes of technology, we emphasize its relational role: how it intervenes in and reshapes interpersonal boundaries between clients and therapists. In contrast to quantitative methods that emphasize hypothesis testing and rely on predefined variables, qualitative approaches are better suited to capturing the complex, multi-layered, and asymmetrical relational dynamics that emerge in therapeutic contexts, along with the subtle dimensions of cognitive collaboration and emotional expression.

Specifically, we adopt Charmaz’s constructivist grounded theory (CGT)~\cite{charmaz_constructing_2012}, which views knowledge as co-constructed between researchers and participants. This approach is particularly well-suited for exploring situated, evolving, and socially embedded phenomena. Rather than aiming for universal generalizations, we seek to understand how meaning, trust, and interpretive alignment are dynamically produced and negotiated within therapist–client interactions, particularly in contexts where AI-mediated tools are introduced as relational mediators.

\subsection{Recruitment Approach and Demographic Overview}

\begin{figure*}[!t]
  \centering
  \includegraphics[width=0.9\textwidth]{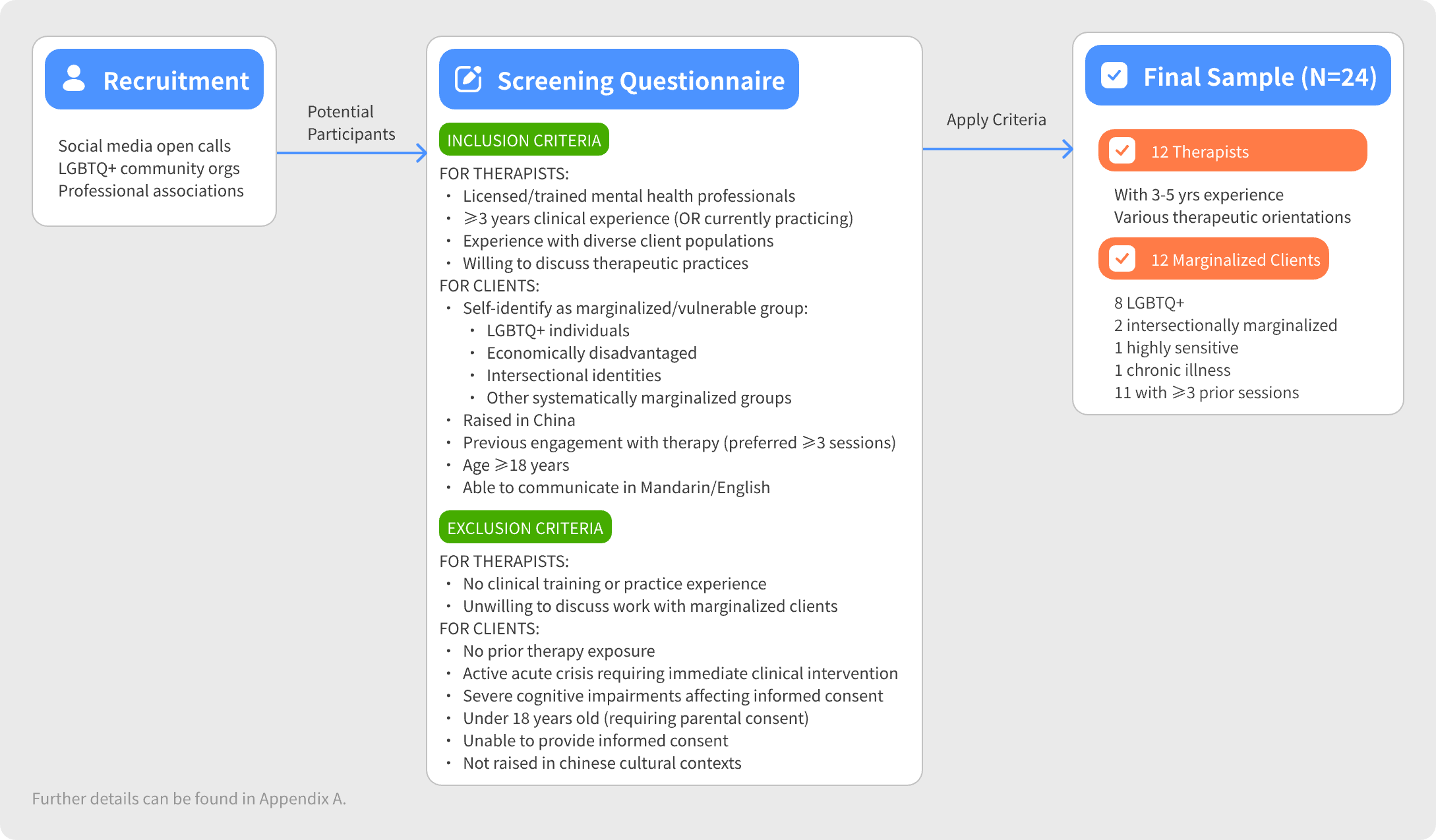}
  \caption{Participant recruitment and screening process. Potential participants were recruited through three channels: social media open calls, LGBTQ+ community organizations, and professional mental health associations. A structured screening questionnaire with inclusion and exclusion criteria was used to identify eligible participants, resulting in a final sample of 24 participants (12 therapists and 12 marginalized clients).}
  \label{fig:recruitment}
\end{figure*}

To capture a range of perspectives, we employed a combination of purposive and snowball sampling strategies. Participants were recruited through three primary channels: (1) open calls disseminated via social media platforms; (2) partnerships with two LGBTQ+ community organizations; and (3) outreach through three professional associations in the mental health field (see Figure~\ref{fig:recruitment}). A total of 24 participants were recruited and evenly divided into two cohorts: therapists (\(n = 12\)) and clients (\(n = 12\)). 

\begin{table*}
\centering
\small
\renewcommand{\arraystretch}{1.4}
\caption{Demographics of Client Participants}
\label{tab:participants-users}
\resizebox{\textwidth}{!}{%
\begin{tabular}{p{0.8cm}p{3.5cm}p{3.5cm}p{6.5cm}p{2.5cm}}
\toprule
\textbf{ID} & \textbf{Therapy Experience} & \textbf{Marginalized Identity} & \textbf{Primary Therapy Modality} & \textbf{Location} \\
\midrule
C1  & $>$20 sessions; $\ge$5 therapists & Highly sensitive & Online counseling platform & Guangdong, CN \\
C2  & 1 walk-in session & LGBTQ+ & School counseling center & Ontario, CA \\
C3  & 11–20 sessions; $\ge$5 therapists & LGBTQ+ & Public-hospital psychiatry; private clinic; online platform & Hong Kong SAR \\
C4  & $>$20 sessions; $\ge$5 therapists & LGBTQ+ & School counseling center; online platform & Beijing, CN \\
C5  & $>$20 sessions; 2–4 therapists & LGBTQ+; full-time housewife; economically dependent & Public-hospital psychiatry; private clinic; online platform & Chongqing, CN \\
C6  & $>$20 sessions; 2–4 therapists & LGBTQ+; economically dependent & Public-hospital psychiatry; private clinic; online \& non-profit services & Beijing, CN \\
C7  & $>$20 sessions; $\ge$5 therapists & LGBTQ+ & Public-hospital psychiatry; school counseling; online platform & Beijing, CN \\
C8  & $>$20 sessions; 2–4 therapists & LGBTQ+ & School counseling center; private clinic & Illinois, US \\
C9  & $>$20 sessions & Chronic illness (not self-identified marginalized) & Life-coaching style sessions & USA \\
C10 & $>$20 sessions; 2–4 therapists & LGBTQ+ & Private clinic; online platform & Kunming, CN \\
C11 & $>$20 sessions; 2 therapists & LGBTQ+ & School counseling center; private clinic & Melbourne, AU \\
C12 & $>$20 sessions; 2 therapists & LGBTQ+ & Public-hospital psychiatry; online platform & London, UK \\
\bottomrule
\end{tabular}}
\end{table*}

For client participants, we prioritized individuals who identified as marginalized or from a vulnerable group, who had prior therapy experience and who were raised in China. We excluded clients who had no prior therapy experience or could not safely participate in a 60–90 minute interview (e.g., active crisis, severe cognitive impairment, or inability to provide informed consent). We also excluded individuals who were not raised in Chinese cultural contexts. 

DeVito et al. describe marginalized groups as those who face structural discrimination and are not well-served by offline counterparts in getting sufficient support \cite{10.1145/3311957.3359442}. Based on that definition, we conceptualize marginalization as structurally shaped, involving systemic barriers, stigmatization, and a lack of equitable access to resources. Following this framing, we include individuals from LGBTQ+ communities, people with lower socioeconomic status, and those living with chronic illness or disability, and so on, because mainstream systems frequently fail to provide appropriate or accessible healthcare. This framing underscores that marginalization is rooted in systemic conditions, not only in one’s identity.

Among the clients, two identified with intersectionally marginalized identities (LGBTQ+ and another marginalized background), eight identified as LGBTQ+, and two as members of other marginalized backgrounds. One client participant (C9) did not self-identify chronic illness as constituting marginalization; we included this case to highlight cultural perspectives on marginalization. All clients were raised in Chinese cultural contexts and eleven had completed at least three prior psychotherapy sessions. We included clients living both in China and abroad to examine how cultural influences shape marginalized identities and therapeutic experiences across settings, and how such influences may evolve through cross-cultural encounters (see Table~\ref{tab:participants-users} for complete demographic details).

For therapist participants, we prioritized licensed or formally trained mental helath professionals with over 3 years of clinical experience or currently practicing, and with experience working with diverse client populations, especially marginalized groups.
 
All therapist participants are Chinese and were trained and currently practice domestically. While not every therapist explicitly stated that they regularly worked with marginalized clients, all had experience serving diverse client populations. In our sampling, we valued therapists' professional expertise as well as their background in supporting marginalized groups.

Table~\ref{tab:participants-therapists} summarizes the professional backgrounds of therapist participants. The therapists represented a range of theoretical orientations, including cognitive-behavioural, humanistic, and psychodynamic, with 3 to 15 years of professional experience across institutional clinics, digital platforms, and private practices. 

\begin{table*}
\centering
\small
\renewcommand{\arraystretch}{1.4}
\caption{Demographics of Therapist Participants}
\label{tab:participants-therapists}
\resizebox{\textwidth}{!}{%
\begin{tabular}{p{0.8cm}p{1.5cm}p{4.2cm}p{3.8cm}p{6cm}}
\toprule
\textbf{ID}& \textbf{Gender}&\textbf{Professional Experience} & \textbf{Marginalized Client Focus} & \textbf{Primary Therapeutic Approaches} \\
\midrule
P1  & Women & Practice since 2022; individual and group counseling & Children, adolescents, LGBTQ+ & Psychodynamic; CBT \\
P2  & Women & Practice since 2019; career-psychology integration & None & CBT; Positive psychology \\
P3  & {Women} & Multiple years in medical center & LGBTQ+ & Integrative (psychodynamic + CBT) \\
P4  & Men & 7 yrs in high school in Tibet; work with blind students & Ethnic minorities; low-income; blind community & Psychoanalysis; CBT; Humanistic \\
P5  & Women & University counseling; 2000+ hrs of practice & None & CBT \\
P6  & Women & University and hospital settings; assessments & Children & Psychodynamic; sandplay; CBT; family therapy \\
P7  & Women & 4 yrs; 500+ hrs & Minority adolescents; LGBTQ+ & Self-psychology; integrative \\
P8  & Women & 5 yrs; PhD-track (Education and Psychology) & Economically disadvantaged students & Humanistic (initially psychoanalysis) \\
P9  & Women & 20 yrs private, corporate, and community work & Low-income; blind community & Communication psychology; transactional analysis; humanistic \\
P10 & Women & 2–3 yrs in university settings & Adolescents & Psychodynamic \\
P11 & Women & $>$2 yrs; full-time school psychologist & Students with limited finances & CBT \\
P12 & Men & Special education schools & Autism and special-needs children & Psychodynamic; Jungian; sandplay; object-relations; client-centered \\
\bottomrule
\end{tabular}}
\end{table*}

\subsection{Data Collection Procedures}

\subsubsection{Semi-Structured In-Depth Interviews}

We conducted semi-structured interviews lasting 60 to 90 minutes with each of the 24 participants. These conversations explored participants' experiences navigating cross-boundary dynamics and identity negotiation in therapy, their reflections on power imbalances in communication and knowledge exchange, and their expectations or concerns regarding the use of LLM-enhanced collaborative mental health systems at different stages of the therapeutic process. We also probed into how emotional expression, trust-building, and self-disclosure unfolded in their interactions. Guided by boundary object theory, we used a flexible interview structure that maintained core themes while allowing participants to shape the direction of the conversation. 

\subsection{Speculative Feature Prototypes: Design Elicitation through Boundary-Informed Probes}
\label{sec:speculative_features}

Recognizing that boundary objects are not static tools but dynamic relational mediators shaped by context, we translated this theoretical perspective into a set of speculative features for LLM-enhanced collaborative mental health systems. These features operationalize our boundary framework in concrete, discussable ways within therapist-client interactions and, guided by empirical insights and existing literature, were introduced as design probes to elicit participants' expectations, relational values, and concerns about trust, autonomy, and alignment, thereby serving as catalysts for discussing the relational implications of such technologies. These twelve features are explicitly designed as dynamic boundary objects that operationalize the three meta-roles of our Dynamic Boundary Mediation Framework: \textit{Epistemic Mediation} (reducing knowledge asymmetries), \textit{Relational Mediation} (rebalancing power dynamics), and \textit{Contextual Mediation} (bridging therapy-life gaps). Each feature embodies boundary object theory's core principles of plasticity (adapting to different therapeutic contexts), coherence (maintaining consistency across social worlds), and mediational agency (actively facilitating cross-boundary collaboration), and rather than functioning as static tools, these features are conceptualized as adaptive socio-technical actors that can shift their primary mediating roles depending on therapeutic stage and relational needs. Below, we outline twelve speculative features utilized in our scenario-based interviews.

\textbf{Epistemic Mediation Features} -- Translating knowledge across therapist-client boundaries:

\begin{itemize}
  \item \textbf{Prescreening}: Functions as an epistemic boundary object that translates clients' subjective experiences into structured information interpretable by therapists, reducing initial knowledge asymmetries while preserving lived experience authenticity~\cite{bendig_next_2022,kruzan_i_2022}.

  \item \textbf{Psychological Education and Self-help Tools}: Acts as a bidirectional knowledge translator, making professional concepts accessible to clients while reducing the "educator burden" on marginalized populations who often must teach therapists about their identities~\cite{liu_using_2022}.

  \item \textbf{Basic Chat Function}: Serves as a low-stakes epistemic boundary space where mutual understanding can develop through informal knowledge exchange before formal therapeutic engagement~\cite{song_typing_2024,fitzpatrick_delivering_2017}.
\end{itemize}

\textbf{Relational Mediation Features} -- Rebalancing power dynamics and fostering relational safety:

\begin{itemize}
  \item \textbf{Privacy Control}: Embodies boundary object plasticity by allowing clients to dynamically negotiate what information crosses the private-professional boundary, redistributing relational power and addressing surveillance concerns among marginalized groups~\cite{stephanidis_privacy_2021}.

  \item \textbf{Empathy Feature}: Creates an intermediate relational space that responds and communicates understanding of individuals' emotional state while maintaining therapeutic coherence across stages, particularly valuable for trust-building with clients who have experienced systemic marginalization ~\cite{de_gennaro_effectiveness_2020}.

  \item \textbf{Therapeutic Journaling}: Functions as a relational boundary object that enables controlled self-disclosure pacing, allowing clients to process emotions in a safe intermediate space before sharing with therapists~\cite{lee_exploring_2021}.
\end{itemize}

\textbf{Contextual Mediation Features} -- Bridging therapy-life boundaries:

\begin{itemize}
  \item \textbf{Between-Session Activities}: Operates as contextual boundary objects that maintain therapeutic continuity while adapting to clients' real-world constraints, particularly important for marginalized clients navigating unsupportive environments~\cite{oewel_approaches_2024}.

  \item \textbf{Crisis Intervention}: Serves as an adaptive boundary mediator that can shift between therapeutic support and real-world crisis management, bridging the gap between clinical safety and lived vulnerability.

  \item \textbf{Mindfulness Exercises}: Functions as portable boundary objects that help translate therapeutic insights into everyday practices, supporting real-world application of clinical learning~\cite{schillings_effects_2024}.
\end{itemize}

\textbf{Cross-Boundary Features} -- Spanning multiple mediational roles:

\begin{itemize}
  \item \textbf{Emotion Tracking}: Acts as a temporal boundary object that creates coherence across episodic interactions while adapting to changing emotional contexts~\cite{duffy_chatbot-based_2024}.

  \item \textbf{Personalization}: Embodies the plasticity principle by allowing the system to adapt its mediating approach to different cultural contexts and communication styles~\cite{kocaballi_personalization_2019}.

  \item \textbf{Multimodal Emotional Expression}: Serves as a translational boundary object that enables expression across different communicative modalities, particularly valuable for clients whose experiences may not easily translate into verbal disclosure~\cite{chu_towards_2024}.
\end{itemize}

These features served as structured stimuli for scenario-based discussions in our interviews. By engaging participants with these probes, we gained nuanced insights into how boundary tensions emerge and how LLM-enhanced collaborative mental health systems might alleviate or exacerbate them across different therapy stages.

\begin{figure*}[!t]
  \centering
  \includegraphics[width=0.55\textwidth]{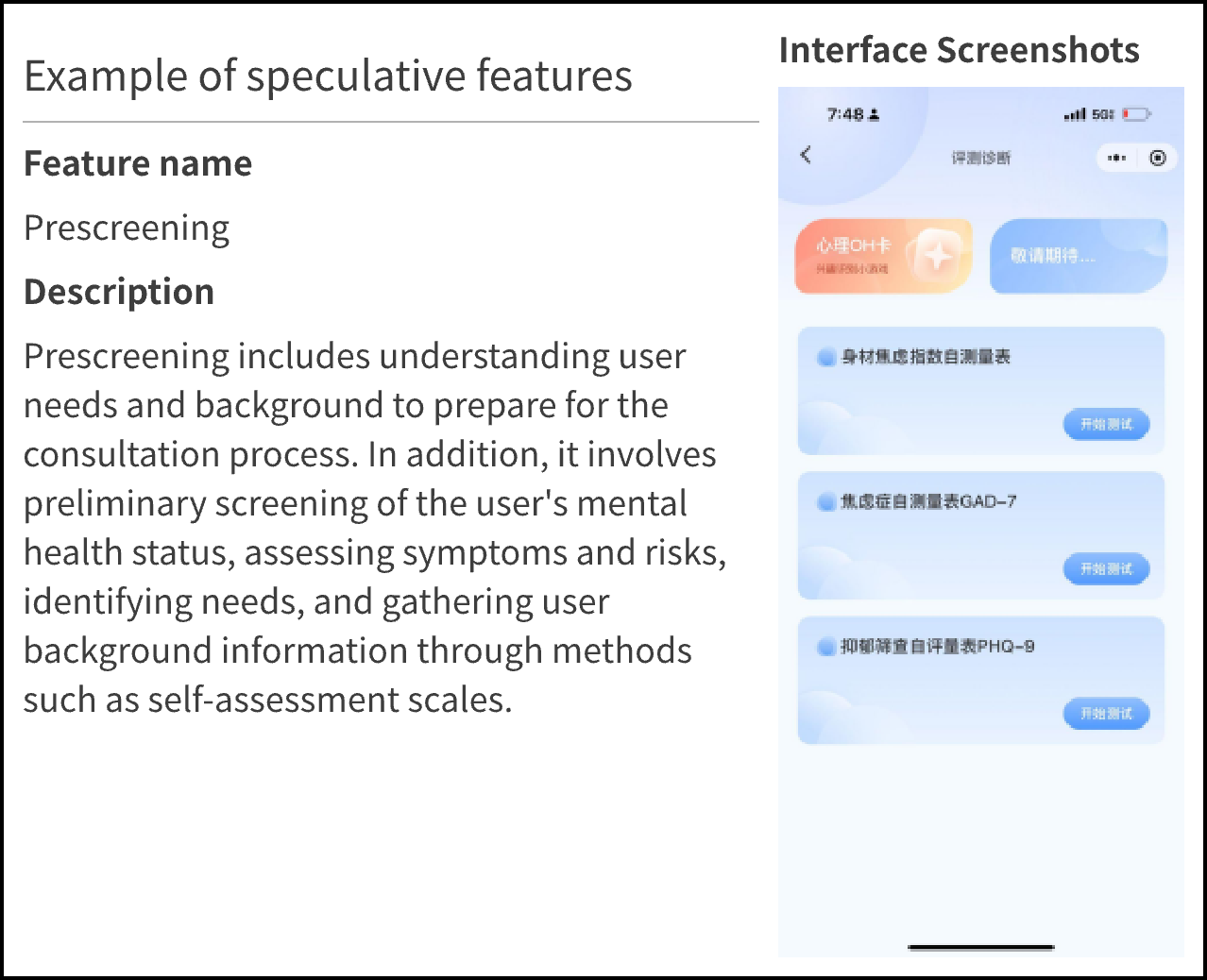}
    \caption{Example of a design probe presented to participants during interviews. Participants 
    were shown annotated interface screenshots for twelve speculative features, including this 
    Prescreening feature, and asked to rate each feature's perceived usefulness (0-10 scale) 
    and provide qualitative reasoning about potential benefits, concerns, and contextual variations.}
    \label{fig:design_probe_example}
\end{figure*}

\subsubsection{Scenario Simulation and Design-Elicitation Probes}

To bridge theoretical abstraction and experiential specificity, we incorporated immersive scenario-based design probes. We presented participants with annotated interface screenshots of the twelve speculative features of LLM-enhanced collaborative systems described above. Figure~\ref{fig:design_probe_example} illustrates one such design probe. Using screen sharing, participants were asked to reflect on these features through quantitative ratings and qualitative reasoning. For each feature, participants first rated its perceived usefulness on a 10-point Likert scale (0=not useful at all; 10=extremely useful), then provided explanations of their ratings, discussing potential benefits, concerns, and contextual variations.

\textbf{Scenario Presentation Protocol:} Each participant was presented with the twelve features in a consistent order to ensure comparability across interviews. Features were presented through annotated wireframe mockups displayed via screen sharing, accompanied by verbal, contextualized explanations from researchers to facilitate participant understanding. For example, the Privacy Control feature was introduced as: "This feature allows you to control what information is stored, shared, or accessed by your therapist, supporting your autonomy and confidentiality. How might this affect your comfort level and willingness to engage in therapy?"

\textbf{Interview Prompting Strategy:} For each feature, we employed a structured questioning approach that explored: (1) Perceived usefulness and concerns; (2) Impact on the therapeutic relationship; (3) Contextual variations across different therapeutic situations. We used follow-up probes to understand participants' underlying reasoning and relational expectations.

\textbf{Stage-Based Evaluation:} Participants were asked to consider each feature's utility across different phases of their therapeutic experience, from initial contact through ongoing care. This temporal perspective helped reveal how feature preferences and concerns might vary depending on therapeutic context and relationship development.

We also adopted the Elaboration Prompting Technique (EPT)~\cite{chi_learning_2001} to encourage deeper interpretive explanation—probing not just feature preferences but also the underlying beliefs, values, and relational expectations associated with each feature. This approach yielded rich contextualized data, illuminating how trust, collaboration, and self-expression were cognitively and emotionally structured in therapeutic interactions.

\subsection{Data Analysis Strategy}

Our analysis followed the principles of constructivist grounded theory and proceeded through three iterative phases.

\textbf{Open Coding:} Two researchers independently conducted line-by-line coding of the interview transcripts, identifying meaningful semantic units related to participants’ therapeutic experiences, such as trust, identity, emotional expression, and interactions with both therapists and AI-based tools.

\textbf{Focused Coding:} Through iterative comparison across cases and therapy stages, we consolidated and refined codes into focused categories. This process allowed us to identify recurring patterns and relational dynamics that shaped participants’ engagement throughout the therapeutic journey.

\textbf{Theoretical Coding:} In the final phase, we integrated these categories into a data-driven conceptual model that highlights how AI-supported systems may mediate therapeutic relationships at different stages of care, leading to the development of our five-stage boundary-negotiation model and the Dynamic Boundary Mediation framework. Throughout the analysis, we employed a constant comparative approach, collaboratively memoing emergent themes and refining interpretations. Discrepancies were resolved through iterative team discussions. The resulting thematic structure can be seen in Appendix (see~\ref{Thematic codebook}).

\subsection{Ethical Considerations}

This study was reviewed and approved by the Institutional Review Board. All data were anonymized through a multi-level de-identification process and securely stored using encryption. Informed consent was obtained from all participants. With informed consent, all interviews were audio-recorded and fully transcribed. They were also informed of their right to withdraw from the study at any time without penalty. Participants could additionally request the deletion of their interview transcripts, without the need to return any compensation.

To ensure accuracy and comfort, all quoted material included in this manuscript was verified with the respective participants. LGBTQ+ participants were provided with enhanced ethical protections, including the option to remain anonymous, skip sensitive questions, and review key findings prior to publication.

All participants received monetary compensation for their time and contributions. Clients were compensated at a rate of 60 RMB per hour, and therapists received 100 RMB per hour. This compensation was intended to recognize their expertise and time. In addition, interviewers were trained in trauma-informed methods and cultural competence. Referrals to professional mental health services were offered within and after the interviews to support participant well-being.

\subsection{Positionality}
As we adopt a constructivist view in conducting this qualitative research, we recognize that our own perspectives shape how we interpret the findings. Most researchers on our team are from China and grew up within Chinese cultural contexts, which informs our understanding of how individuals might perceive mental health in these settings. A few researchers are from LGBTQ+ communities or are long-term allies, which sensitizes us to issues faced by marginalized identity groups. Our team also includes members from different disciplines, such as psychology, human–computer interaction, and health informatics. In addition, one researcher has therapy experience in both the U.S and China, which also shapes our view of mental health practices across different cultural contexts.
\section{Findings}

Our empirical investigation sought to answer two central research questions. First, we explored the challenges and dynamics shaping trust, self-disclosure, and communication between therapists and marginalized clients in China across various therapeutic stages (RQ1). Second, we examined how LLM-enhanced collaborative mental health systems can support and improve the quality of these relationships throughout these stages (RQ2).

\subsection{RQ1: Challenges and Dynamics Shaping Trust, Self-Disclosure, and Communication in Therapy with Marginalized Clients}

\begin{figure}  
  \centering
  \includegraphics[width=0.85\linewidth]{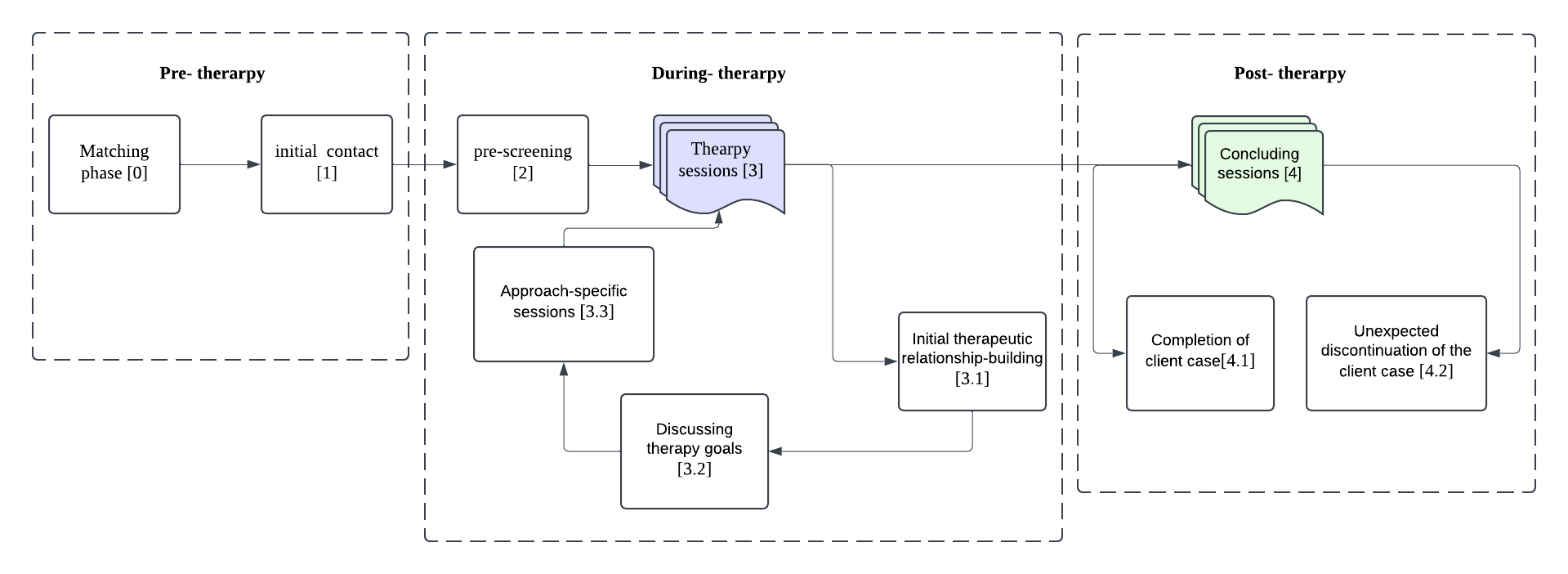}  
  \caption{Overview of the typical multi-stage psychological counselling service flow, from matching to completion, as derived from participant interviews. This process view provides context for the subsequent five-stage model of relational boundary negotiation.}
  \label{fig:therapy_service_flow}
\end{figure}

To address our first research question, which examines the challenges and dynamics that influence trust, self-disclosure, and communication between therapists and clients from marginalized groups during different stages of therapy, we analyzed interviews with twelve psychotherapists and twelve marginalized clients in China, including those from LGBTQ+ communities, economically disadvantaged backgrounds, and those living with chronic illness or high sensitivity. The analysis shows that the therapeutic process involves continuous \emph{boundary work}, including the negotiation of meaning, professional practices, and authority. These relational elements are shaped by how such negotiations unfold. As illustrated in Figure~\ref{fig:therapy_service_flow}, this process typically follows a multi-stage service flow.

From this analysis, we identified five key stages where relational boundary negotiation becomes especially important: Matching and Initial Contact, Initial Trust-Building and Self-Disclosure, Client Expression, Between-Session Continuity, and Closure and Real-World Integration. The following subsections detail the specific challenges observed at each stage.

\subsubsection{Stage 1: Matching and Initial Contact — Foundational Challenges to Trust and Communication}
The initial stage of matching and contact immediately surfaces challenges to establishing trust and open communication. Divergent understandings of "therapeutic expertise" create an \textit{epistemic boundary}. While therapists often foreground formal credentials, marginalized clients prioritize lived experience and identity fluency. As C8 emphasized, "What we need is not some certificate from a multicultural training seminar, but a therapist who actually understands our lived realities." This misalignment can make the initial match emotionally taxing and undermine early trust, as institutional knowledge clashes with community-based needs for situated understanding. Standardized intake processes often exacerbate this by flattening identity, hindering effective communication of needs and a fair assessment of compatibility.

\subsubsection{Stage 2: Initial Trust-Building and Self-Disclosure — Navigating Relational Boundaries and Vulnerability}
Once contact is made, the dynamics of trust-building and initial self-disclosure are paramount. Trust is not instant but a "relational structure that gradually unfolded"(P7). Premature pushing by therapists or a lack of professionalism can rupture this fragile trust (C3: "It is hard to find someone with a strong sense of professional ethics."). Clear, ethical boundaries are crucial for facilitating disclosure. However, ambiguous confidentiality, especially within institutional settings (Therapist: "I work at a university… many students don't trust their counselors..."), poses significant challenges to disclosure and trust.

\subsubsection{Stage 3: Client Expression (Who Explains to Whom?) — Asymmetrical Communication and the Burden of Disclosure}
A key dynamic shaping communication and disclosure is the "burden of client expression." Clients, particularly from marginalized groups, often spend significant time educating therapists about their identities or cultural contexts (C1: “The therapist didn't really understand my upbringing...”; C11: “A new one came in, and I had to retell my story all over again.”). This "having to teach the therapist" (as therapists like P7 acknowledge the need for interpretation with some clients) places an asymmetrical labor on the client. While potentially part of disclosure, this repetitive "boundary labor" due to therapist turnover or lack of initial understanding can exhaust clients, thereby eroding trust in the therapeutic process and inhibiting deeper, more meaningful communication and self-disclosure.

\subsubsection{Stage 4: Between-Session Continuity — Misaligned Expectations for Communication and Support}
The period between formal sessions reveals misalignments in expectations around communication and relational continuity, affecting perceived trust and support. While therapists maintain professional temporal boundaries (P2: "Clients might send us messages, but we don't necessarily have to respond..."), marginalized clients often seek more continuous support and an "ongoing presence"(C1). This points to a need for mechanisms that mediate these different expectations for communication and presence.

\subsubsection{Stage 5: Therapeutic Closure and Real-World Integration — Challenges in Transferring Trust and Communication Skills}

The final stage highlights the difficulty of transferring the trust, communication patterns, and self-disclosure practices developed within therapy into unsupportive real-world contexts. Clients often encounter a stark contrast between the relative safety of therapeutic settings and the social risks present in their daily lives(C5). P4 similarly reflected that the “relatively safe space” of therapy is disconnected from everyday social environments, where different power relations and communicative expectations dominate. These conditions make it challenging for clients to maintain the relational skills and emotional openness cultivated during therapy once they re-enter less supportive or stigmatizing settings.

\subsection{RQ2: LLM-Enhanced Collaborative Mental Health Systems: Supporting and Improving Therapy Quality Across Therapeutic Stages}

Building on our analysis of the key challenges and dynamics shaping trust, self-disclosure, and communication throughout the therapeutic process (RQ1), we now address our second research question: How can LLM-enhanced collaborative mental health systems support and improve therapy quality across different stages of therapy? We posit that these systems, by functioning as dynamic boundary objects~\cite{star_institutional_1989}, can offer unique affordances to mediate these relational elements. They can adopt functions such as facilitating understanding between parties, creating safe intermediate spaces, or providing structured forums for negotiation throughout the various stages of the therapeutic process.The following subsections detail how these systems can support each therapeutic stage.

\subsubsection{Stage 1 (Matching \& Initial Contact): Fostering Initial Trust and Transparent Communication}
LLM-enhanced collaborative mental health systems can improve the quality of initial therapeutic interactions by supporting clearer communication of needs and expectations, helping to establish a stronger foundation for trust.

\paragraph{Enhancing Transparency and Informed Matching.} To reduce mismatches in expertise, systems can make therapist information more transparent and help clients express what they are looking for, especially regarding identity. In addition, pre-screening tools can support clients in understanding their own needs and communicating their emotional state more clearly to potential therapists as mentioned by two clients (C12, C2), therefore it will further improve the quality of the initial connection.

\paragraph{Creating a Safer Initial Space for Disclosure.} Participants expressed a fear of being judged in traditional therapeutic settings. As one participant noted, “I actually feel that communicating with GPT carries less risk than therapists” (C1). LLM-based systems may help reduce this perceived risk by offering a less intimidating environment for clients to express their needs and vulnerabilities. By lowering initial barriers to disclosure, such systems can encourage clients to articulate their concerns more openly from the start.

\subsubsection{Stage 2 (Initial Trust-Building \& Self-Disclosure): Scaffolding Trust and Enabling Gradual, Controlled Disclosure}
These systems can improve relationship quality by providing a safer, more controlled environment for the delicate processes of trust-building and initial self-disclosure.

\paragraph{Cultivating a Consistent and Non-Judgmental Relational Space.} For marginalized clients who are particularly sensitive to judgment (e.g. C1, C6), the consistent and nonjudgmental nature of an LLM-enhanced collaborative mental health system can serve as a reliable point of contact. This sense of stability can help foster safety.

\paragraph{Empowering Clients with Control over Disclosure.} Concerns about data misuse were raised by both therapists and clients, including mishandling sensitive information (C2, C6) and unintended social consequences (C12). To address these concerns, participants emphasized the need for systems to provide fine-grained privacy controls and the desire for selective disclosure. Similarly, P5 expressed that being able to share information at their own pace would make them feel more in control and safer when opening up. Providing such controls can enhance trust and emotional safety, thereby improving the quality of self-disclosure in therapeutic interactions.

\subsubsection{Stage 3 (Client Expression): Facilitating Clearer Communication and Reducing Disclosure Burden}
In this stage, we found that LLM-enhanced collaborative mental health systems can improve the quality of communication and disclosure by reducing the client's "educator burden."

\paragraph{Mediating Understanding of Marginalized Experiences.} To alleviate the burden of "having to teach the therapist" (C1, C11, C6, C5), systems could provide (with consent) therapists with curated, identity-affirming information or help clients articulate complex experiences (P7), thus making in-session communication more focused on therapeutic work rather than basic education. 
\paragraph{Supporting Culturally Sensitive Dialogue.} By potentially assisting therapists in accessing culturally relevant knowledge (P1, P4) or helping clients phrase sensitive topics, these systems can foster more attuned and respectful communication, strengthening the therapeutic alliance and trust.

\subsubsection{Stage 4 (Between-Session Continuity): Enhancing Relational Connectedness and Consistent Communication}
In stage 4, these systems can enhance the sense of relational continuity by offering consistent, mediated channels for ongoing communication and support.

\paragraph{Providing Asynchronous Support and Maintaining Connection.} Balancing clients’ desire for a sense of ongoing presence (C1) with the practical and emotional boundaries of therapists (C6,P2), LLM-enhanced collaborative mental health systems can offer support through features such as "24/7 AI companionship" (P2) and asynchronous communication channels including asynchronous messaging, mood tracking, and reflective journaling (C5, C9). This can enhance feelings of being supported and connected, improving the perceived quality of the therapeutic relationship between sessions. Tools for reminders and tracking (P2, P1, P11, P4) sustain an ongoing communicative link to the therapeutic process.

\paragraph{Improving Emotional Attunement in System Responses.} Participants like C5 and C6 described a “robotic feeling” in chatbot interactions, where responses felt scripted, impersonal, or emotionally flat. To foster stronger relational engagement, systems need to move beyond generic replies and instead demonstrate emotional sensitivity and continuity.
\subsubsection{Stage 5 (Therapeutic Closure \& Real-World Integration): Supporting the Sustained Quality of Self-Management and Communication}

In this final stage, we found that LLM-enhanced collaborative mental health systems can support clients in extending therapeutic gains beyond the boundaries of formal sessions. These systems help bridge the gap between the safe, structured environment of therapy and the often less supportive, more complex realities of everyday life by reinforcing communication skills, self-awareness, and emotional regulation.

\paragraph{Bridging Therapeutic Learning to Real-World Communication.} To address the “limited contextual portability” of in-session progress noted by C5 and P4, LLM-enhanced systems can offer interactive tools that promote reflection and self-awareness (C11). Clients can revisit therapeutic communication strategies and practice applying appropriate self-disclosure boundaries in real-world situations. 

\paragraph{Ensuring Continuous Support through Crisis Intervention.} The need for post-therapy crisis support was emphasized by P4 and P5, who underscored the importance of accessible assistance during emotionally vulnerable moments. P6 and C2 pointed out that accurate and timely detection of crises is critical, while P11 highlighted the value of connecting users to immediate and appropriate sources of help. LLM-enhanced systems that provide real-time responsiveness and support pathways can strengthen users’ sense of security and trust, thereby contributing to the continuity of care beyond formal therapy.
\section{Discussion}

Our findings illuminate the nuanced relational landscape of psychotherapy involving marginalized clients in China, revealing distinct challenges and dynamics that shape trust, self-disclosure, and communication across five therapeutic stages (RQ1). Furthermore, our exploration into LLM-enhanced collaborative mental health systems (RQ2) highlighted their potential to act as dynamic boundary objects, adaptively mediating these crucial relational elements. This section synthesizes these insights. We first formally introduce and elaborate on the \textbf{Dynamic Boundary Mediation Framework}, a conceptual model emerging from our stage-based analysis of both the challenges and the mediational opportunities. We then discuss this framework's broader implications for AI-mediated care and translate its core tenets into actionable design guidelines for developing relationally accountable AI that is particularly attuned to the needs of marginalized populations.
\begin{figure}
  \centering
  \includegraphics[width=0.95\linewidth]{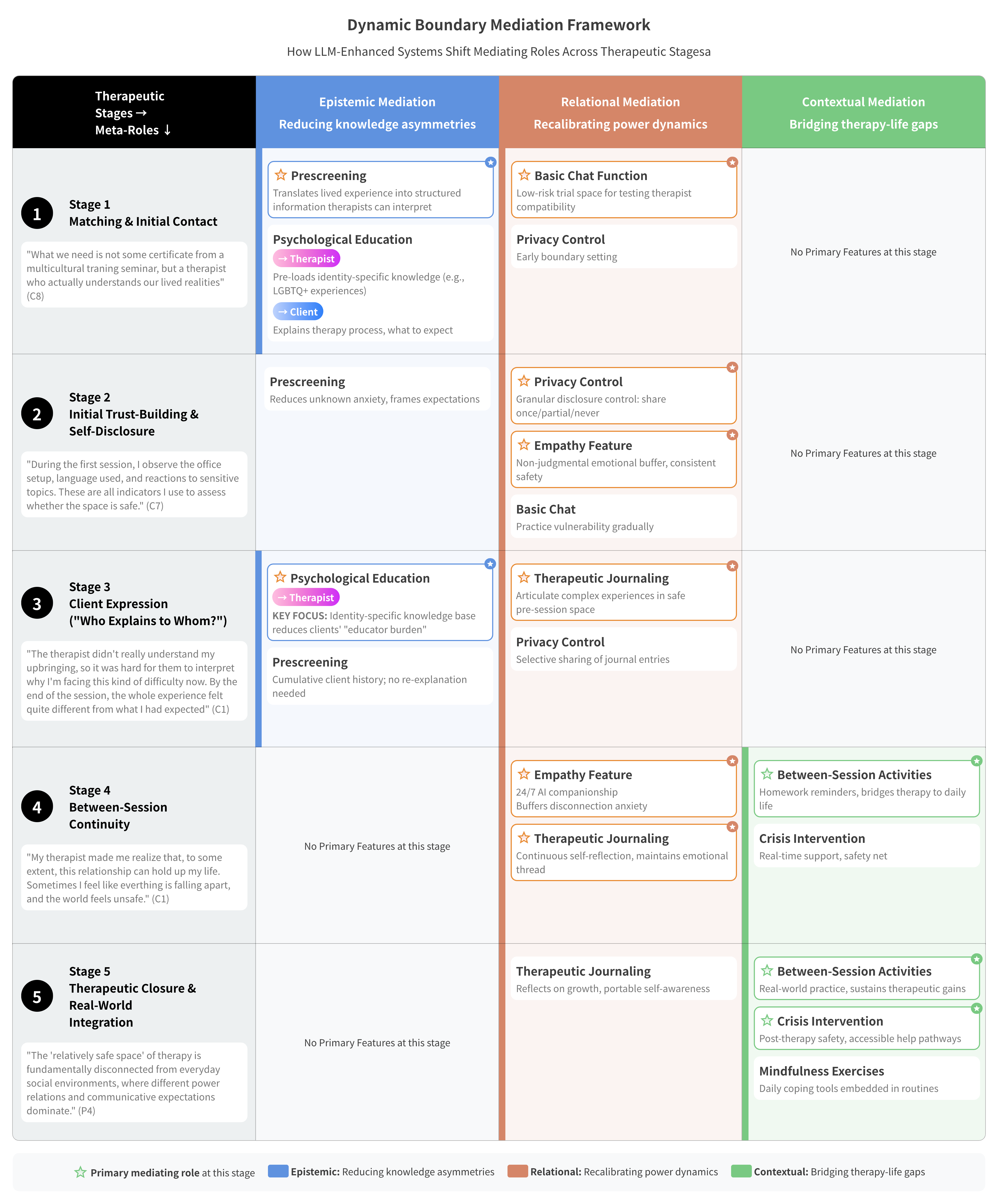}
    \caption{The Dynamic Boundary Mediation Framework illustrating how LLM-enhanced collaborative mental health systems shift mediating roles across five therapeutic stages. The framework highlights three meta-roles: Epistemic Mediation (reducing knowledge asymmetries), Relational Mediation (recalibrating power dynamics), and Contextual Mediation (bridging therapy-life gaps). Specific features are mapped to each stage based on empirical findings from interviews with marginalized clients and therapists in China. Starred features indicate primary mediating functions at each stage.}
  \label{fig:five_stage_model}
\end{figure}

\subsection{The Dynamic Boundary Mediation Framework: Theorizing LLM-Enhanced Systems as Adaptive, Stage-Aware, and Community-Informed Relational Agents}

The preceding stage-based analysis of how LLM-enhanced collaborative mental health systems can address relational tensions (RQ2) reveals their capacity to function as dynamic boundary objects whose mediating roles can and should evolve throughout the therapeutic journey. Synthesizing these observations, we propose the \textbf{Dynamic Boundary Mediation Framework}. This framework conceptualizes these systems not merely as tools, but as agentive socio-technical mediators that build connection between clients and therapists. This framework actively transform therapeutic relationships by using their boundary, spanning functions to respond to shifting relational needs and contextual demands, particularly those of marginalized clients, at each specific stage identified in our RQ1 findings. This framework builds upon the twelve speculative features from Section~\ref{sec:speculative_features}, demonstrating how boundary object theory applies to digital therapeutic contexts.


The framework highlights three key meta-roles and dimensions through which these systems perform their mediational work, with varying emphasis across the therapeutic stages:

\begin{enumerate}
    \item \textbf{Epistemic Mediation:} Reducing knowledge asymmetries by translating marginalized experiences, cultural contexts, and identity-specific concerns into formats that enhance mutual understanding. This is operationalized through features such as \textit{Prescreening} (which translates subjective experiences into structured therapeutic information) and \textit{Psychological Education} (which reduces the "educator burden" on marginalized clients). This mediation is crucial during \textit{Stage 1 (Matching \& Initial Contact)} to address misaligned expertise and in \textit{Stage 3 (Client Expression)} to alleviate asymmetrical knowledge labor.
    
    \item \textbf{Relational Mediation:} Rebalancing power dynamics and fostering relational safety by creating intermediate spaces where clients can exert greater control over interaction pace, disclosure depth, and communication style. Features like \textit{Privacy Control} (enabling dynamic boundary negotiation) and \textit{Empathy Feature} (creating buffered relational space) exemplify this mediational work. This is vital in \textit{Stage 2 (Initial Trust-Building \& Self-Disclosure)} for scaffolding trust and in \textit{Stage 4 (Between-Session Continuity)} for enhancing connectedness. It is also important in Stage 1 (reducing early uncertainty), Stage 3 (supporting balanced emotional expression), and Stage 5 (maintaining relational grounding during therapeutic closure and real-world integration).
    
    \item \textbf{Contextual Mediation:} Bridging the therapy-life gap by helping translate clinical insights into culturally appropriate, real-world practices. This is embodied in features such as \textit{Between-Session Activities} (maintaining therapeutic continuity in daily life) and \textit{Crisis Intervention} (bridging clinical support with real-world vulnerability). This mediation is key in \textit{Stage 4 (Between-Session Continuity)}  \textit{Stage 5 (Closure \& Real-World Integration)}, especially for marginalized clients navigating unsupportive environments.
\end{enumerate}

This framework extends boundary object theory~\cite{star_institutional_1989} by emphasizing how these systems \textit{dynamically shift} mediating functions across therapeutic stages, moving beyond static interoperability to active relational transformation. It underscores their potential to redistribute emotional labor and enhance client agency.

However, realizing the full potential of these mediational roles is challenged by what participants termed the "robotic feeling", defined as interactions perceived as artificial and lacking genuine warmth. As C5 explained, "When talking to a chatbot, you say a lot, and then it just replies with 'we understand you'… that's not appealing to me." This undermines the system's ability to foster relational safety or emotional continuity. Participants noted the lack of relational continuity as a major weakness, with conversations staying narrowly focused without connecting to past discussions (C6). This contrasts with human therapists' ability to follow clients' emotional journeys and make them feel truly seen (P4). While participants acknowledged AI's utility for immediate support (C11), they voiced concerns about handling complex, layered emotional experiences. These insights underscore that LLM-enhanced systems must transcend scripted responses and develop capabilities for genuine emotional sensitivity, robust relational memory, and adaptive pacing to foster a sense of being truly understood and relationally held.

These empirical challenges highlight important tensions between boundary object theory's assumptions and digital implementation realities. While traditional boundary objects gain their mediational power through human interpretation and collaborative meaning-making, the speculative features we designed must achieve similar effects through algorithmic responses. 

A crucial aspect further enriching both Epistemic and Relational Mediation, particularly for marginalized clients, is how trust navigation transcends formal credentials and often relies on \textit{community-based validation}. Our findings show these clients often rely on peer networks and community endorsements as adaptive responses to systemic mistrust and past negative experiences within formal systems. This contrasts with some therapists' views on universal rapport-building (P3: "I just treat them like any other client…"), highlighting a potential disconnect. For LLM-enhanced collaborative mental health systems to effectively function as mediators, the Dynamic Boundary Mediation Framework thus posits they must be \textit{community-informed}: designed to recognize, respect, and potentially integrate these informal trust pathways, for example, by signposting to community-vetted resources, thereby aligning more closely with how marginalized clients already navigate and establish relational safety.

This insight directly informed our design of features like \textit{Prescreening} and \textit{Basic Chat Function}, which could potentially integrate community validation mechanisms, and \textit{Privacy Control}, which acknowledges marginalized clients' sophisticated strategies for managing institutional trust. However, the challenge remains: how can algorithmic systems authentically represent and respect these informal trust networks without co-opting or undermining them?

This dynamic, stage-aware, and community-informed perspective, while cognizant of current technological limitations in achieving deep relationality and genuine emotional attunement, offers a novel lens for understanding and designing AI in sensitive care contexts, prioritizing relational accountability.

\subsection{Broader Implications: Towards Relationally Accountable AI in Diverse Care Contexts}

Although this study focuses on mental health care in China, the Dynamic Boundary Mediation Framework provides broader insights for AI-supported care in other domains, such as education, social services, and legal counseling. These settings often share similar characteristics, including power imbalances, knowledge asymmetries, cultural tensions, and the need to build trust, especially when working with vulnerable or marginalized populations.

The central claim of this framework is that the effectiveness of AI in sensitive care contexts depends not only on its technical capacity but also on its ability to mediate relational dynamics in ways that are adaptive, stage-aware, context-sensitive, and emotionally responsive. This perspective challenges traditional approaches in human-computer interaction that emphasize single-purpose functionality. Instead, it suggests that AI systems should be designed to shift roles based on the evolving needs of users and care relationships. These roles may include supporting knowledge translation, facilitating emotional expression, or anchoring interactions in specific cultural or institutional contexts.

This research also highlights an important design choice: AI systems can either reinforce existing inequalities or help transform them. Tools that are generic and insensitive to local trust mechanisms or emotional needs may unintentionally deepen exclusion. In contrast, systems designed as dynamic and community-informed boundary mediators have the potential to redistribute emotional and cognitive burdens, create safer spaces for self-expression, and better connect institutional systems with users’ lived experiences.

As AI technologies continue to be adopted in high-stakes care environments, it becomes essential to consider their capacity for mediating relational boundaries. Designing AI that is relationally accountable and structurally responsive is crucial for ensuring that technological innovation promotes equity and social justice, particularly for those most affected by systemic marginalization.

\subsection{From Framework to Practice: Design Guidelines for Relationally Accountable and Marginalization-Attuned AI Systems}

Translating the Dynamic Boundary Mediation Framework and our empirical insights (including the challenges of the "robotic feeling" and the importance of community validation) into actionable guidance, we propose five specific design guidelines (DGs). These guidelines aim to help developers, researchers, and policy makers operationalize relational accountability, ensuring LLM-enhanced mental health systems function as effective, equitable, and \textit{marginalization-attuned} boundary mediators:

\begin{enumerate}[label=\textbf{DG\arabic*}, leftmargin=*, itemsep=1em]

    \item \textbf{Stage-Aware, Community-Informed, and Emotionally Attuned Role Shifting}: Design systems to dynamically adjust their mediating functions across the five therapeutic stages. In early stages (matching and initial contact), prioritize culturally sensitive communication, trust-building through community validation mechanisms, and clarifying expectations using identity-relevant knowledge. During mid-stages (trust-building, client expression, and between-session continuity), focus on supporting gradual, controlled disclosure, alleviating the "educator burden" by facilitating clear articulation of experiences, maintaining relational continuity through contextual memory of past interactions, and delivering emotionally responsive engagement that moves beyond superficial validation. In the late stage (Therapeutic closure and real-world integration), emphasize prompts for maintaining therapeutic gains, cues for applying insights to daily life, and ongoing support, with all elements underpinned by a flexible privacy framework that adapts to evolving disclosure needs throughout the therapeutic journey.

    \item \textbf{Negotiable Data Visibility within a Flexible Privacy Architecture}: 
    Implement granular, multi-layer consent mechanisms (\textit{share once}, \textit{share partially}, \textit{keep private}) for data shared with or recorded by the system. This empowers users, especially marginalized clients concerned about surveillance or stigma, to dynamically modulate data visibility to professionals or for longitudinal modeling, acknowledging that trust and disclosure comfort evolve across stages. Ensure transparent data governance and clearly communicate how shared data might influence the AI's relational memory and responses.

    \item \textbf{Contextualized Relational Memory and Deep Adaptive Emotional Attunement}: 
    Employ distilled "relational summaries" (themes, triggers, progress from earlier stages, significant emotional expressions) rather than full transcripts to maintain narrative coherence and continuity. This must be coupled with interaction designs that naturally surface relevant history and AI responses that demonstrate deep adaptive emotional attunement—striving to "genuinely see the pain, the longing, the hope" (P4) by moving beyond scripted empathy to reflect understanding of the client's unique emotional journey, communication style, and evolving needs. This mitigates re-identification risks, reduces the “panopticon effect,” and directly counters the "Robotic feeling" by fostering coherent and genuinely supportive interactions over time.

    \item \textbf{Community-Validated Onboarding and Proactive, Dynamic Identity-Sensitive Knowledge Integration}: 
    Particularly in early stages (Matching), integrate community validation by embedding endorsements from, or links to, trusted peer organisations and advocacy groups. Concurrently, proactively incorporate identity-sensitive knowledge about diverse marginalized communities, not as static databases, but as a dynamic capacity for the system to learn and adapt its understanding of identity-salient concepts and cultural nuances, thereby reducing clients' "educator burden" and enhancing therapist attunement. Provide clear pathways for redress and user feedback on the system's cultural and identity competence.

    \item \textbf{Pervasive, Nuanced Context-Adaptive Empathy to Overcome Robotic Feeling}: 
    Across all interaction stages, design for context-adaptive empathy that actively works to overcome the "robotic feeling." Couple foundation models with capabilities to understand and respond to nuanced cues (e.g., dialect, honorifics, culturally specific expressions relevant to marginalized groups, shifts in emotional tone). Dynamically vary response length, formality, and emotional tenor according to user state, expressed needs, cultural norms, and the specific therapeutic stage. The goal is for interactions to consistently feel appropriately empathetic, validating, respectful, and genuinely responsive, rather than like an exchange with a machine merely processing input (C5).
\end{enumerate}

Collectively, these guidelines aim to redistribute emotional and cognitive labor, foreground user agency, and ensure that LLM-enhanced collaborative mental health systems act as equitable, community-informed, and stage-aware boundary mediators, rather than as amplifiers of existing social and power hierarchies. By embedding these considerations into system design, especially the pursuit of genuine emotional attunement and respect for community trust mechanisms, developers can move beyond purely functional success metrics toward creating technologies that are relationally just, structurally responsive, and contextually grounded, particularly for those navigating systemic discrimination.
\section{Limitations and Future Work}
While this study offers valuable insights into how LLM-enhanced collaborative mental health systems can strengthen therapist–client relationships, several limitations constrain the generalizability of our findings. Our participant sample was limited in geographic and demographic diversity, with most LGBTQ+ participants from urban, first-tier cities in China with high education levels and digital literacy. The findings may not reflect experiences of individuals in rural or under-resourced areas where stigma is stronger and access limited. The study did not include other vulnerable populations such as older adults, or ethnic minorities, limiting generalizability. Cultural diversity within China across regions, dialects, and ethnic communities was not fully represented. Given the sensitivity of topics involving sexual orientation, gender identity, and trauma, participants may have withheld or softened responses despite confidentiality assurances, potentially limiting data depth and authenticity. The feature evaluation relied on subjective feedback and expert judgment outside real-world settings, limiting validity. Participants' responses may have been shaped by their understanding of AI technologies rather than direct prototype engagement. 

Despite these limitations specific to our Chinese sample, the core theoretical contributions of the Dynamic Boundary Mediation Framework have broader applicability. The three meta-roles of Epistemic Mediation, Relational Mediation, and Contextual Mediation address fundamental challenges in therapeutic relationships that transcend cultural boundaries: knowledge asymmetries between professionals and clients exist across healthcare systems globally, power imbalances are inherent to institutional care relationships, and the therapy-life gap affects clients regardless of cultural background. However, specific manifestations of these challenges likely vary across cultural contexts. According to Hofstede’s cultural dimensions\cite{hofstede_dimensionalizing_2011}, collectivist countries like China tend to have high power distance and may therefore place strong value on professional credentials. However, our empirical findings show that marginalized groups such as the LGBTQ+ community in China often value community validation instead. This points to a more nuanced view of Chinese cultural dynamics, and suggests that similar patterns may exist in other cultural contexts as well,with different cultural norms interacting and manifesting at different stages of the therapy experience. Future research should account for these variations through more inclusive sampling, culturally adaptive design, and investigation of AI usage in actual therapeutic settings.

\section{Conclusion}
This study reconceptualizes LLM-enhanced collaborative mental health systems as dynamic relational mediators that actively transform therapeutic relationships, particularly for marginalized clients facing social and structural disadvantages. Our interviews with marginalized clients and therapists in China revealed trust and communication ruptures rooted in cultural mismatch and emotional labor asymmetries, which these systems can address through culturally attuned and privacy-aware interventions. We derive design principles for marginalized users that emphasize identity-sensitive matching, flexible disclosure pacing, and community validation. These systems must recognize users as relational agents and build trust through cultural sensitivity rather than standardization. Our findings underscore the value of culturally situated approaches, with participants emphasizing peer-validated recommendations, indirect communication respect, and disclosure control. These expectations reflect deeply social orientations that system design must honor. The dynamic mediation framework has broader implications for care-oriented fields like education, legal support, and social work, offering a pathway to designing AI that strengthens relational dimensions of care. Ethical safeguards ensuring privacy, transparency, and cultural respect are essential, particularly where therapy records have broader life implications. We advocate for AI systems designed as dynamic boundary mediators that foster safety, rebalance relational labor, and honor clients' lived experiences through culturally and emotionally attuned design.
\section{Acknowledgments}
This work is supported by the Natural Science Foundation of China (Grant No. 62302252).
\bibliographystyle{ACM-Reference-Format}
\bibliography{ref_updated}

\appendix
\section{Appendix}

\subsection{Screening Questionnaire for Interview Recruitment}

    \textbf{Title:} Designing and Developing a Mental Health Chatbot: Exploring Innovations in Mental Health Services
    
    \textbf{Introduction:}  
    Have you ever participated in psychological counseling, or are you interested in innovations in mental health services?  
    We are researchers working on designing and developing a mental health chatbot to improve accessibility and personalization in mental health care. Your experiences and insights are vital to our study.
    
    \textbf{Participation:}  
    After completing this questionnaire, if your background aligns with the target user group of our study, our research team may contact you for a follow-up interview. If not, we appreciate your time and support, though we will not be able to follow up individually.
    
    \textbf{Incentives:}  
    Participants will receive monetary compensation of 60 RMB per hour. An additional 30 RMB will be provided for each extra half hour.
    
    \textbf{Privacy Statement:}  
    Participation is entirely voluntary. You may withdraw at any time without providing a reason. Upon request, we will delete any data you have submitted.  
    All data will be anonymized to ensure your identity cannot be determined, and raw data will be destroyed at the end of the study. Non-identifiable data may be used for academic publications and shared datasets.
    
    \textbf{Contact Information:}  
    If you have questions about the study or data privacy, please contact the research team (email address removed for peer review).  
    \textit{This study has received ethical approval from an institutional review board.}
    
    \subsection*{Screening Questions}
    
    \begin{enumerate}
      \item \textbf{What is your age?}  
      \begin{itemize}
        \item Under 18  
        \item 18–24  
        \item 25–34  
        \item 35–44  
        \item 45–54  
        \item 55 and above
      \end{itemize}
    
      \item \textbf{Do you identify with any of the following minority/vulnerable groups?} (Select all that apply)  
      \begin{itemize}
        \item LGBTQ+  
        \item People with disabilities  
        \item Adolescents (under 18)  
        \item Economically disadvantaged  
        \item Individuals with ADHD or ASD  
        \item Pregnant women / Stay-at-home mothers / Single mothers / Low-income or financially dependent women  
        \item Ethnic minorities  
        \item Other: \underline{\hspace{4cm}}
      \end{itemize}
    
      \item \textbf{Have you received psychological counseling?}  
      \begin{itemize}
        \item Never  
        \item Previously  
        \item Currently receiving counseling
      \end{itemize}
    
      \item \textbf{If yes, how many sessions have you had in total?}  
      \begin{itemize}
        \item 1–5  
        \item 6–10  
        \item 11–20  
        \item More than 20
      \end{itemize}
    
      \item \textbf{Where have you received counseling?} (Select all that apply)  
      \begin{itemize}
        \item School counseling center  
        \item Public hospital psychiatric department  
        \item Private clinic / psychologist  
        \item Online counseling platform  
        \item Other: \underline{\hspace{4cm}}
      \end{itemize}
    
      \item \textbf{How many different therapists have you worked with?}  
      \begin{itemize}
        \item 1  
        \item 2–4  
        \item 5 or more
      \end{itemize}
    \end{enumerate}

\clearpage

\clearpage
\section{Interview Protocol for Mental Health Professionals}

This semi-structured interview protocol is designed to investigate the practices, challenges, and perspectives of licensed mental health professionals, particularly in relation to their therapeutic workflows, experiences with marginalized populations, and views on AI-supported mental health tools. The goal is to inform the design of human–AI collaborative systems that enhance accessibility, empathy, and trust in mental healthcare.

All interviews will be audio-recorded and transcribed. Personally identifiable information will be removed during transcription and data anonymization. Selected anonymized excerpts may be cited in academic publications. Participation is entirely voluntary. Interviewees may skip any question or withdraw from the interview at any time without explanation. Each session will last approximately 60 minutes, with adjustments depending on the flow of conversation. Compensation will be provided upon completion.

\subsubsection*{Section A: Professional Background and Clinical Context}
\begin{itemize}
  \item Could you briefly describe your educational and professional background (e.g., training, licensure, clinical setting, years of practice)?
  \item What types of clients do you primarily serve (e.g., students, working adults, pregnant women)?
  \item Have you worked with populations considered socially marginalized or structurally vulnerable (e.g., LGBTQ+ individuals, people with disabilities, adolescents, individuals with chronic illnesses)?
\end{itemize}

\subsubsection*{Section B: Experiences with Vulnerable Populations}
\begin{itemize}
  \item What mental health challenges are most commonly reported within these populations?
  \item What barriers have you encountered in establishing therapeutic relationships with these clients?
  \item What strategies do you use to establish and maintain trust in therapy, particularly with clients who have experienced discrimination or trauma?
  \item How do you adapt your therapeutic approach to meet the specific needs of these groups?
  \item Do you refer clients to external resources or support networks (e.g., community organizations, digital tools, peer-led groups)? If so, how do you evaluate their relevance and reliability?
\end{itemize}

\subsubsection*{Section C: Therapeutic Frameworks, Processes, and Cultural Adaptation}
\begin{itemize}
  \item What is your primary theoretical orientation (e.g., CBT, psychodynamic, person-centered)? Why did you choose this framework?
  \item Given that many psychological frameworks are Western in origin, how do you adapt your methods to align with the cultural norms and expectations of clients in your local context?
  \item Could you describe your typical therapeutic workflow—from intake to termination?
  \item How many sessions are usually involved in a full treatment cycle, and what does each phase typically include?
  \item Which stage of therapy tends to be the most resource-intensive for you, and why?
  \item What tools or assessments do you use to evaluate clients’ progress or mental health status?
  \item How do you document client sessions? Are your records standardized or personalized? What type of information do they typically include?
  \item Have you encountered situations where clients were reluctant to share sensitive information due to privacy concerns? How did you navigate such scenarios?
  \item Have you used any collaborative digital tools or electronic health systems as part of your therapeutic practice? If yes, how have they influenced your workflow?
\end{itemize}

\subsubsection*{Section D: Perceptions of Mental Health Chatbots}
\begin{itemize}
  \item Are you familiar with any AI-based mental health tools or conversational agents (e.g., Woebot, XiaoTian, ChatGPT, Pi)?
  \item Based on your experience or knowledge, what potential benefits do you see in using such tools for mental health care?
  \item What limitations or risks do you foresee, particularly regarding clinical effectiveness, safety, or data protection?
  \item How do you perceive the level of acceptance of such technologies among your peers or within your professional community?
  \item Do you believe such chatbots could complement your clinical practice? Why or why not?
    \begin{itemize}
      \item If yes: What specific roles or functionalities (e.g., screening, journaling, session check-ins, therapy homework) would be most beneficial to your workflow?
      \item If no: What theoretical, ethical, or technical limitations do you consider insurmountable?
    \end{itemize}
  \item If such tools were to be integrated into the therapeutic process, what types of data or interactions between clients and the chatbot would be most useful for your practice?
  \item In your view, can these systems meaningfully contribute to sustaining therapeutic gains after therapy ends? Why or why not?
  \item Do you believe these tools can effectively serve clients across different cultural backgrounds? What types of cultural adaptations would be necessary?
  \item Which populations or use cases do you think are most appropriate for chatbot-based mental health tools (e.g., self-guided users, those with mild to moderate symptoms)?
\end{itemize}

\subsubsection*{Section E: Additional Reflections}
\begin{itemize}
  \item In your clinical experience, have you encountered clients who use metaphysical or spiritual practices (e.g., astrology, tarot) to address emotional distress? How do you interpret or respond to these practices in therapy?
\end{itemize}

\section{Interview Protocol for Clients}

This semi-structured interview protocol is designed to explore the lived experiences, emotional needs, and mental health support practices of individuals who have engaged with therapy or other psychological services. Particular attention is given to clients from vulnerable or marginalized backgrounds, as well as their perceptions of AI-based mental health chatbots. The goal is to inform the development of inclusive and ethically sound mental health technologies.

All interviews will be audio-recorded and transcribed. Personally identifiable information will be removed during transcription and data anonymization. Selected anonymized excerpts may be cited in academic publications. Participation is entirely voluntary. Interviewees may skip any question or withdraw from the interview at any time without explanation. Each session will last approximately 60 minutes, with adjustments depending on the flow of conversation. Compensation will be provided upon completion.

\textbf{Interview Objectives}
\begin{itemize}
  \item Understand your mental health journey and therapy experiences.
  \item Explore challenges, trust dynamics, and disclosure practices during therapy.
  \item Examine the influence of cultural, familial, and social systems on therapy.
  \item Gather perspectives on the role, design, and limitations of mental health chatbots.
  \item Identify concerns related to data privacy, trust, and the use of AI in mental health contexts.
\end{itemize}

\textbf{Opening Question}  
To begin, could you briefly introduce yourself (e.g., age range, profession, region)? If you are comfortable, you may also share whether you identify as part of a minority or vulnerable group. This information will help contextualize your experiences in our analysis.

\vspace{1em}
\textbf{Section A: Mental Health Journey and General Background}
\begin{itemize}
  \item Can you describe how your mental health journey began? What led you to seek help?
  \item What kinds of methods or resources have you used (e.g., therapy, self-help, online content)?
  \item Have you searched for mental health-related content on the internet or social media? What kinds of content were you looking for? Were they helpful?
  \item Have you ever used AI-based tools such as ChatGPT, Wenxin Yiyan, or Doubao to explore emotional or mental health issues?
\end{itemize}

\vspace{1em}
\textbf{Section B: Experiences with Therapy}
\begin{itemize}
  \item Have you participated in psychological counseling or therapy? If so, could you briefly describe your experience?
  \item How did you find your therapist(s)? What was the process like?
  \item Have you ever changed therapists? What motivated that decision?
  \item How would you describe your relationship with your therapist? What worked and what didn’t?
  \item At which points during the therapy process did you face challenges (e.g., finding a therapist, opening up, ending therapy)?
  \item Do you feel your therapist was able to empathize with you? What made you feel understood (or not)?
  \item How equal or hierarchical do you perceive the therapist–client relationship to be? Did you feel pressured, judged, or dismissed?
  \item Did your identity (e.g., gender, sexual orientation, ethnicity) affect your interactions with your therapist?
  \item Did you ever feel your therapist made assumptions or showed bias toward your background?
  \item Do you believe your therapist understood the cultural or societal context of your experiences?
\end{itemize}

\vspace{1em}
\textbf{Section C: Trust and Self-Disclosure}
\begin{itemize}
  \item Did you trust your therapist? What helped build that trust?
  \item Were there moments when your trust in the therapist weakened? What caused it?
  \item How did you decide what to share with your therapist? Did you disclose personal matters early on or gradually?
  \item What factors made it easier or harder to open up in therapy?
  \item Would you be willing to disclose private or sensitive topics to a mental health chatbot? Why or why not?
\end{itemize}

\vspace{1em}
\textbf{Section D: Stakeholders and Social Ecosystems}
\begin{itemize}
  \item Were there other people involved in your therapy journey (e.g., family members, school counselors, teachers)?
  \item What roles did they play? Did they support or interfere with your mental health efforts?
  \item Did you feel any conflict of interest between different parties (e.g., parent–therapist–client)?
  \item How did you navigate your own rights, responsibilities, and privacy in such situations?
\end{itemize}

\vspace{1em}
\textbf{Section E: Social Support and Expressive Modalities}
\begin{itemize}
  \item Have you had access to peer support, online communities, or emotional support groups?
  \item What role did these networks play in your coping process?
  \item What are your views on using alternative or creative modes (e.g., art, music, metaphysical practices) for emotional expression in therapy?
  \item Have you ever experienced playful or less formal therapeutic moments (e.g., games, storytelling)? Did they help?
\end{itemize}

\vspace{1em}
\textbf{Section F: Mental Health Chatbots and AI-based Support Tools}
\begin{itemize}
  \item Are you familiar with any mental health chatbots or AI-based mental health tools (e.g., Emohaa, Woebot, XiaoTian)?
  \item Have you used any of them? If yes, how was the experience?
  \item Compared to a human therapist, do you share more or less with an AI assistant? Why?
  \item What features would you expect from a trustworthy mental health chatbot?
  \item Do you believe such tools can support emotional needs? Build empathy? Sustain therapeutic outcomes?
  \item Would you want your therapist to have access to your chatbot interactions? Why or why not?
  \item How should such tools handle privacy, data protection, and crisis situations?
  \item Under what conditions would you prefer using a chatbot over a human counselor?
  \item What concerns, if any, do you have about using AI for mental health?
  \item Would you be open to long-term use of such tools? What would influence your willingness?
\end{itemize}

\vspace{1em}
\textbf{Section G: Reflection and Closing}
\begin{itemize}
  \item Based on your overall experience, what role do you think therapists should play?
  \item If you have had multiple therapists, what traits or approaches made some more effective than others?
  \item Is there anything else you wish to share about your therapy journey or your views on AI in mental health?
\end{itemize}

\section{Thematic Framework on Mental Health Experiences and Technology}
\label{Thematic codebook}
\subsection{Thematic Framework among Clients}
\begin{longtable}{p{0.25\textwidth} p{0.1\textwidth} p{0.6\textwidth}}
\caption{Codebook for Thematic Analysis of Interviews (Clients)} \\
\toprule
\textbf{Theme} & \textbf{Code} & \textbf{Description} \\
\midrule
\endfirsthead

\multicolumn{3}{c}%
{\tablename\ \thetable\ -- \textit{Continued from previous page}} \\
\toprule
\textbf{Theme} & \textbf{Code} & \textbf{Description} \\
\midrule
\endhead

\midrule \multicolumn{3}{r}{\textit{Continued on next page}} \\
\endfoot

\bottomrule
\endlastfoot

Theme 1: Vulnerable Groups and Mental Health & 1 & Current problems \\
& 1.1 & Financial challenges \\
& 1.2 & Societal pressure \\
& 1.3 & Invisibility of LGBTQ+ Experiences in Mainstream Therapy \\
& 1.4 & Generational and Cultural Barriers in Mental Health Understanding \\
& 2 & Mental health needs \\
& 3 & Experiences of Vulnerable Groups in China \\
& 3.1 & Limited resources/access to mental health services \\
& 4 & Identity exploration \\
& 4.1 & Distinction Between Identity Acceptance and Coming Out \\
& 4.2 & Social support networks as resources \\
& 5 & Blurred line between mental health help with other resources \\
& 6 & Traumatic experience \\
& 6.1 & Acute health event triggered help seeking \\
& 7 & Social Media Resources \\
& 8 & Navigating Power Dynamics in Mental Health Decisions \\

Theme 2: Therapy Experience & 1 & Trust \\
& 1.1 & Building trust through flexible message response patterns and negotiation \\
& 1.2 & Building Trust Through Recommendations from Friends and Community Networks \\
& 1.3 & Building trust through shared values \\
& 2 & Journey of seeking mental health help \\
& 2.1 & Episodic, Situational, and Pragmatic Therapy Engagement \\
& 2.2 & How vulnerable find therapists \\
& 2.3 & Holistic, systemic, and transformative therapy \\
& 2.4 & Conclude session \\
& 3 & Expectation \\
& 3.1 & Expectations of therapy \\
& 4 & Preference on therapists and psychological schools of thought \\
& 5 & Privacy \\
& 6 & Emotional support \& Self-disclosure \\
& 7 & Problem-solving \\
& 8 & Empathy \\
& 9 & Reflection on therapy experience \\
& 10 & Problems encountered during therapy \\
& 11 & Discomfort \\
& 12 & Biases \\
& 13 & Challenges \\
& 14 & Quality of the therapy \\
& 15 & Overall Experience with the therapy \\
& 16 & Power dynamics between therapists and clients \\
& 17 & Crisis intervention \\
& 18 & In-person therapy benefits: non-verbal communication in therapy \\
& 19 & Relationship \\

Theme 3: Mental Health Chatbot & 1 & Privacy \\
& 2 & Safety \\
& 3 & Effectiveness \\
& 4 & Concerns \\
& 5 & Expectations \\
& 6 & Trust \\
& 7 & Empathy \\
& 8 & Emotional support \\
& 9 & Understanding of the technology \\
& 10 & Interesting use case \\
& 11 & Dislikes \\
& 12 & Relationship with chatbot \\
& 13 & Accessibility \\
& 14 & Design opportunities \\
\end{longtable}

\subsection{Thematic Framework among Mental Health Professionals}
\begin{longtable}{p{0.25\textwidth} p{0.1\textwidth} p{0.6\textwidth}}
\caption{Codebook for Thematic Analysis of Interviews (Therapists)} \\
\toprule
\textbf{Theme} & \textbf{Code} & \textbf{Description} \\
\midrule
\endfirsthead

\multicolumn{3}{c}%
{\tablename\ \thetable\ -- \textit{Continued from previous page}} \\
\toprule
\textbf{Theme} & \textbf{Code} & \textbf{Description} \\
\midrule
\endhead

\midrule \multicolumn{3}{r}{\textit{Continued on next page}} \\
\endfoot

\bottomrule
\endlastfoot

Theme 1: Chatbot and Care Delivery & 1 & Access to care (e.g., time, stigma, geography) \\
& 1.1 & 24/7 availability and convenience \\
& 2 & Quality of care \\
& 2.1 & Emotional support \\
& 3 & Therapist–client relationship \\
& 3.1 & Engagement and interaction \\
& 3.2 & Trust \\
& 4 & Self-disclosure facilitation \\
& 5 & Privacy \\
& 5.1 & Patient consent, data security, risk of data misuse \\
& 8 & Integration into therapy workflow \\
& 8.1 & Therapist-side integration \\
& 8.2 & Client-side integration \\
& 9 & Limitations of chatbot use \\
& 9.1 & Use and concerns with psychological scales \\
& 10 & Crisis intervention and responsibility \\
& 11 & Perceptions of therapy \\
& 12 & Ethical concerns (e.g., privacy, human-AI boundary, job replacement) \\
& 13 & Emotion data tracking and analysis \\
& 14 & Current tech use in therapy \\
& 15 & Design functions/features \\
& 16 & Design opportunities (e.g., empathy, multimodal input, memory, cultural adaptation) \\
& 17 & General user flow (one therapy session) \\
& 18 & Number of consultations per client \\
& 19 & Emergent/Interesting insights \\

Theme 2: Clients and Vulnerable Groups & 1 & Holistic view of mental health problems \\
& 2 & Cultural and societal considerations \\
& 2.1 & Observations of cultural influence \\
& 2.2 & Language and communication adaptation \\
& 2.3 & Universal strategies vs cultural specificity \\
& 2.4 & Localization and adaptation of methods \\
& 3 & Needs of vulnerable groups \\
& 3.1 & Challenges (e.g., stigma, bias, therapist discomfort) \\
& 3.2 & Support systems and structures \\
& 3.3 & Client–therapist matching and sensitivity \\
& 4 & Empathy and emotional attunement \\
& 5 & Awareness and cognition \\
& 6 & Trust-building techniques in therapy \\
& 7 & Professionalism and boundaries \\
& 8 & Balancing responsibilities across stakeholders (e.g., school, family) \\
& 10 & Client avoidance and self-protection mechanisms \\
& 11 & Common mental health challenges (e.g., relationships, emotions) \\
& 12 & Problem-solving roles of therapists \\
& 13 & Alternative methods beyond formal psychotherapy \\
\end{longtable}

\section{Scenario-Based Feature Rating Task for Mental Health Chatbot Evaluation}

As part of the semi-structured interview protocol, participants completed a scenario-based feature evaluation task. They were asked to imagine interacting with a comprehensive AI-based mental health chatbot powered by a large language model. The chatbot was described as supporting users throughout the entire therapeutic process—before, during, and after sessions—by providing features such as journaling, emotion tracking, psychoeducational content, and privacy controls. The system was positioned as a collaborative tool for enhancing communication, engagement, and emotional support between clients and therapists.

Participants were instructed to evaluate each feature on a scale from 0 (not useful at all) to 10 (extremely useful), and to briefly explain their ratings. The aim was to capture both quantitative assessments and the underlying personal, emotional, or relational reasoning behind their evaluations.

\textbf{Scenario Prompt:}
Imagine you are using an all-in-one AI mental health chatbot. This tool can be used before, during, and after therapy sessions. It helps you engage with psychological support, reflect on your emotions, track your mental state, and access relevant resources. For each of the following features, please rate how useful it would be to you (0–10), and briefly explain your reasoning.

\begin{itemize}
\item \textbf{Prescreening}
Collects background information, presenting concerns, and emotional status through self-report questionnaires and screening tools. Designed to match users with appropriate services or prepare for therapeutic engagement.

\item \textbf{Basic Chat Functionality}
Enables casual interaction with the chatbot, including daily check-ins, emotional updates, and informal conversations.

\item \textbf{Empathy Feature}
Simulates emotionally supportive and human-like responses to foster a sense of care and connection.

\item \textbf{Privacy Feature}
Allows users to control what information is stored, shared, or accessed—supporting autonomy and confidentiality.

\item \textbf{Therapy Activity – Journaling}
Provides a space for expressive writing to support self-reflection, emotional processing, and regulation.

\item \textbf{Between-Session Therapy Activities}
Offers structured prompts or exercises to maintain therapeutic momentum between sessions.

\item \textbf{Psychology Educational Content}
Delivers accessible information on mental health topics, coping strategies, and emotional literacy.

\item \textbf{Self-Help Tools}
Includes guided exercises such as deep breathing, progressive muscle relaxation, or grounding techniques for independent emotion regulation.

\item \textbf{Emotion Tracking}
Allows users to log and visualize emotional patterns over time to support mood awareness and trigger identification.

\item \textbf{Mindfulness Exercises}
Offers guided present-focused practices to reduce anxiety, stress, or emotional overwhelm.

\item \textbf{Crisis Intervention}
Provides immediate coping strategies and referral options during emotional crises or high-distress situations.

\item \textbf{Personalization}
Tailors language, recommendations, and feature delivery based on user preferences, communication style, or therapeutic goals.
\end{itemize}

\end{document}